\documentclass[twoside,english]{article}
\usepackage[T1]{fontenc}
\usepackage[latin9]{inputenc}
\usepackage{geometry}
\usepackage{float}
\usepackage{amsmath}
\usepackage{graphicx}
\usepackage{amssymb}
\usepackage{epstopdf}

\makeatletter
\newcommand{\lyxaddress}[1]{
\par {\raggedright #1
\vspace{1.4em}
\noindent\par}
}

\makeatother

\usepackage{babel}

\begin{document}

\title{The direct scattering study of the parametrically driven nonlinear
Schr\"odinger equation}

\author{CP Olivier$^{1,2}$ and NV Alexeeva$^{2}$}

\maketitle

\lyxaddress{$^{1}$South African National Space Agency, 1 Hospital Street, Hermanus, South Africa 7200 Email: Dr.Carel.Olivier@gmail.com\\
$^{2}$Department of Mathematics and Applied Mathematics, University
of Cape Town, Private Bag Rondebosch 7701, South Africa. Email: Nora.Alexeeva@uct.ac.za}
\begin{abstract}
The term {}``direct scattering study'' refers to the calculation
and analysis of the discrete eigenvalues of the associated Zakharov-Shabat
(ZS) eigenvalue problem. The direct scattering study was applied to
time-dependent oscillating solitons that arise as attractors in the
parametrically driven nonlinear Schr\"odinger equation. Four different types of 
attractors within the parameter space are identified, each with a unique soliton
content structure. These structures include radiation-induced
nonlinear modes and soliton complex structures. The different types of attractors 
are used to characterise the dependence of the attractors on damping and driving
parameters. Period-doubling bifurcations are shown to affect the radiation emissions
of oscillating solitons. The role of soliton complex structures and 
radiation in the formation of spatio-temporal chaos is also identified.
\end{abstract}

\section{Introduction}

The development of inverse scattering theory \cite{key-1} -- \cite{key-5}
led to the realization that the scattering data can be used as nonlinear
modes to analyse solitons in nonintegrable systems. Kaup \cite{key-6}
pioneered the adiabatic analysis, a method that determines the effect
of small perturbations on solitons. The method was generalized to
the perturbed inverse scattering method \cite{key-7,key-8} so that
the effects of radiation could be taken into account. These methods
led to insight into the effect of many perturbations, including
linear and nonlinear dissipative and forcing terms. The restriction
of these methods is that they can only be applied to known solutions
of the unperturbed equation. One way to overcome this difficulty is
to calculate the direct scattering data numerically. We refer to this
method as the direct scattering study.

The direct scattering study was introduced by Overman and co-workers
\cite{key-9} who developed a numerical scheme to calculate the scattering
data for periodic potentials. They applied this scheme to investigate
chaotic solitons that arise in the damped-driven sine-Gordon equation.
Subsequently Ablowitz and co-workers \cite{key-10,key-11} applied
this scheme to analyse chaotic solutions that were observed in the
integrable NLS equation for periodic potentials. Their study led to
the identification of homoclinic crossings, the result of numerical
errors, as the source chaos. Bofetta and Osborne \cite{key-39} developed
a numerical scheme for the calculation of the scattering data for
decaying potentials defined on the real line. Their scheme was used
to study soliton generation in the integrable NLS equation from Gaussian
initial potentials \cite{key-12}, chirped potentials \cite{key-13}
and disordered optical fields propagating in nonlinear bulk waveguides
\cite{key-14}. It was also applied to study solitons in many nonintegrable
NLS equations where the inclusion of various effects were studied,
including linear damping \cite{key-18,key-19}, periodic amplifiers
\cite{key-15}, higher-order dissipation combined with linear gain
\cite{key-16}, quintic nonlinearities \cite{key-20} and many more
\cite{key-17} -- \cite{key-24}.

In this paper we illustrate the ability of the direct scattering study
to analyse soliton dynamics in two different ways. Firstly the identification
of different soliton structures can lead to insight into the dynamical
structure and spatial features of solitons in nonlinear equations.
Secondly the ability to calculate energy-related measures of radiation
allows one to analyse the effect of radiation in the dynamics. For
this purpose we consider a model that exhibits a rich diversity of
soliton solution, namely the parametrically driven nonlinear Schr\"odinger
(PDNLS) equation, given by \begin{equation}
i\phi_{t}+\phi_{xx}+2\left|\phi\right|^{2}\phi-\phi=h\phi^{*}-i\gamma\phi.\label{eq:PDNLS}\end{equation}
Here $h>0$ corresponds to the driving strength, while $\gamma>0$
represents the damping coefficient. The PDNLS equation is an important
model for resonant phenomena in hydrodynamics \cite{key-25} -- \cite{key-27}
nonlinear optics \cite{key-28} -- \cite{key-30} and many more (see \cite{key-31,key-32}
and the references therein).

The PDNLS equation admits a trivial zero solution. \cite{key-34}
The zero solution is stable whenever the driving strength $h\leq h_{c}$,
where \[
h_{c}=\sqrt{1+\gamma^{2}}.\]
When the driving strength exceeds the damping coefficient $\gamma\leq h$,
the PDNLS equation has two additional time-independent soliton solutions
\cite{key-34} \begin{equation}
\psi^{\pm}\left(x\right)=A^{\pm}e^{-i\theta^{\pm}}\mbox{sech}\left(A^{\pm}x\right),\label{eq:PDNLS solitons}\end{equation}
where $A^{\pm}=\left(1\pm\sqrt{h^{2}-\gamma^{2}}\right)^{1/2},$ $\theta^{+}=\frac{1}{2}\mbox{arcsin}\left(\gamma/h\right)$
and $\theta^{-}=\theta^{+}-\frac{\pi}{2}.$ The solution $\psi^{-}$
is unstable for all choices of damping and driving strengths $\gamma$
and $h$ respectively. The $\psi^{+}$ solution has a stable region.
For the large damping regime $\gamma\geq0.378$, the $\psi^{+}$ solution
is stable whenever $\gamma\leq h\leq h_{c}$. In the smaller damping
regime $\gamma<0.378$ the $\psi^{+}$ solution destabilizes due to
a Hopf bifurcation. In this regime the stability criteria for $\psi^{+}$
is given by $\gamma\leq h<h_{H},$ where $h_{H}<h_{c}$ is the critical
driving strength where the Hopf bifurcation occurs. The Hopf bifurcation
gives rise to time-dependent soliton solutions. We therefore only
consider the small damping regime.

\begin{figure}
\begin{centering}
\includegraphics[scale=0.6]{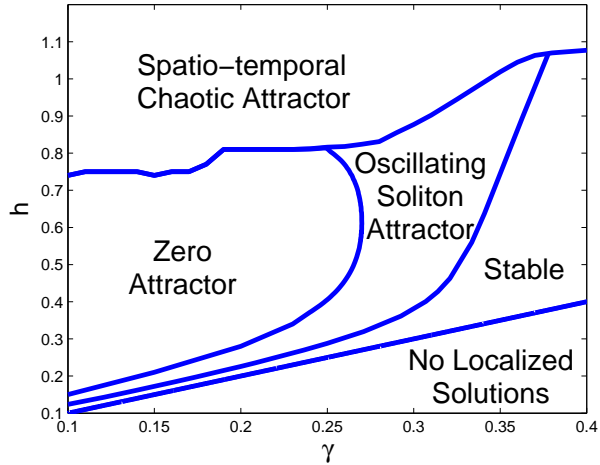}
\par\end{centering}

\caption{\emph{Attractor chart of the $\psi^{+}$ solution. }\label{fig:Attractor chart}}
\end{figure}

Bondila and co-workers \cite{key-33} generated the time-dependent
soliton attractors through direct simulation. They also reported two
other attractors in the unstable region of $\psi^{+}$, namely the
zero attractor and spatio-temporal chaotic attractors. These results
are summarized in Figure~\ref{fig:Attractor chart}. They also characterized
the temporal structure of the soliton attractors, reporting that the
majority of soliton attractors have a simple 1-period mode of oscillation.
The only exception is in a narrow band near the zero attractor region
where period-doubling bifurcations lead to temporally chaotic solitons.
Barashenkov and co-workers \cite{key-34} showed that these chaotic
soliton attractors are caused by homoclinic instabilities.

From a spatial point of view, soliton attractors consist of a soliton
part and a radiation part. The soliton part consists of a quiescent
(non-propagating) soliton that oscillates temporally. The radiation
part consists of waves with small amplitudes and widths relative to
the soliton part of the solutions. These waves arise symmetrically
about the soliton, and propagate away from the soliton. Radiation
waves are subject to dissipation associated with the linearized equation,
and account for losses in the system. Time-dependent soliton attractors
arise due to a balance between energy gained through driving and energy
lost through dissipation \cite{key-32}.

The interaction between solitons and radiation plays an important
role in the dynamics of the PDNLS equation. Alexeeva and co-workers
\cite{key-40} showed that radiation is responsible for instabilities
in the undamped case. Shchesnovich and Barashenkov \cite{key-37} used
inverse scattering theory to construct a reduced system of ODE's using
a single nonlinear mode and a single radiation mode. Their system
produced the period-doubling route to chaos, as well as the zero attractor
region. However they were unable to reproduce the full complexity
of the attractor chart Figure~\ref{fig:Attractor chart}. Barashenkov
and co-workers \cite{key-32} used a linearized analysis to study the
effect of damping on radiation emissions.  Their analysis show that 
smaller damping strength suppresses the emission
of radiation waves. It should be noted that this analysis
ignores the effects of interaction between solitons and radiation.

In this paper we use the direct scattering study to gain insight into
the role that soliton structures, radiation waves and their interaction
play in the dynamics of the PDNLS equation. The paper is structured
as follows: In Section 2 we discuss the generation of soliton attractors
of the PDNLS equation. We analyse these results by calculating basic
characteristics associated with these attractors. We show the difficulties
of separating radiation from the soliton by inspection. In Section
3 we apply the direct scattering study to the solitons of the PDNLS
equation. We reveal four different types of soliton attractors, based
on the soliton content. The results show that large radiation emissions
can excite additional nonlinear modes. We also reveal soliton complexes
in the form of breather-like structures that may result from large
driving strengths. Based on these results we describe the effect of
damping and driving strengths on the structure of the soliton and
its radiation emissions. In Section 4 we consider the period-doubling
route to temporal chaos and its effect on the radiation emissions
of the soliton. Here radiation emission measurement is used to describe
the effect of period-doubling bifurcations on radiation emissions.
We also consider soliton transients within the zero attractor region,
and identify the role of solitons and radiation in the annihilation
of the soliton transient. In Section 5 we consider the spatio-temporally
chaotic region. We identify the formation of multiple solitons as
the source of chaos formation, and discuss the role of soliton complex
structures and radiation in this process. In Section 6 we conclude
with a brief discussion of the results.

\section{Basic characteristics of soliton attractors}

To generate the soliton attractors we follow the direct simulation
approach of Bondila and co-workers \cite{key-33}. This is done by
integrating the unstable $\psi^{+}$ solution \eqref{eq:PDNLS solitons}
numerically. Numerical errors provide destabilizing perturbations
that repulse the orbit, forcing it to a different solution, or attractor,
after a sufficiently long interval of integration. One could interpret
this as constructing a heteroclinic orbit that approaches $\psi^{+}$
and the attractor as $t\rightarrow-\infty$ and $t\rightarrow\infty$
respectively. We used a fourth-order pseudospectral split-step scheme
for integration with an interval length $L=200$, $N=2048$ grid points
and a step size $\Delta t=0.005$. Direct simulation is ideal for
constructing quasi-periodic and temporally chaotic solutions. It also
allows one to calculate soliton transients in regions where no soliton
attractors exist.

In this section we use the numerically generated solutions to study
the effects of damping and driving on the oscillating soliton attractors.
For 1-period soliton attractors, the temporal period $T$ of these
attractors depends on both the damping and driving strengths. For
a fixed driving strength, an increase in damping strength results
in a decrease in the temporal period of the resulting soliton attractor.
Likewise, for a fixed damping strength, an increase in driving strength
leads to a decrease in the temporal period of the resulting soliton
attractor. Mathematically this means that the temporal period of 1-period
attractors can be written as a function $T\left(\gamma,h\right)$
that satisfies\[
\frac{\partial T}{\partial\gamma}<0\mbox{ and }\frac{\partial T}{\partial h}<0.\]

Both the soliton and radiation parts of the soliton attractors also
depend on the damping and driving strengths. Lets first consider the
soliton part of the solution. One of the most important properties
of the soliton is its amplitude, defined as\begin{equation}
A_{\gamma,h}\left(t\right)=\left|\psi_{\gamma,h}\left(0,t\right)\right|.\label{eq:Amplitude}\end{equation}
Here $\psi_{\gamma,h}\left(x,t\right)$ corresponds to the soliton
attractor that arises for damping and driving strengths $\gamma$
and $h$ respectively. We define the magnitude of temporal oscillation
in terms of the soliton amplitude, given by\begin{equation}
m\left(\gamma,h\right)=\mbox{max}\left\{ A_{\gamma,h}\left(t\right)\right\} -\mbox{min}\left\{ A_{\gamma,h}\left(t\right)\right\} .\label{eq:Magnitude of oscillation}\end{equation}
The magnitude of temporal oscillation reflects the change that the
amplitude undergoes during each temporal oscillation. For a fixed
damping strength, the magnitude of temporal oscillation increases
when the driving strength is increased, i.e.\[
\frac{\partial m}{\partial h}>0.\]
However, for a fixed driving strength, an increase in the damping
strength leads to a decrease in the magnitude of temporal oscillations,
i.e.\[
\frac{\partial m}{\partial\gamma}<0\]

The damping and driving have a similar effect on radiation emissions
of the soliton attractors. To measure the radiation we use the radiation
emission amplitude defined as\[
E_{\gamma,h}=\mbox{max}\left\{ R_{\gamma,h}\left(t\right)\right\} ,\]
where $R_{\gamma,h}\left(t\right)=\psi_{\gamma,h}^{(m)}\left(x_{0},t\right)-\psi_{\gamma,h}^{(m)}\left(x_{1},t\right),$
$\psi_{\gamma,h}^{(m)}\left(x,t\right)=\mbox{Im}\left\{ \psi_{\gamma,h}\left(x,t\right)\right\} ,$
and $x_{0}\left(t\right)>0$ and $x_{1}\left(t\right)>0$ are the
closest points to the origin where $\psi_{\gamma,h}^{(m)}$ has a
local maximum and local minimum respectively.The radiation emission
amplitude indicates the size of radiation waves that is emitted during
each temporal period. For a fixed damping strength, an increase in
the driving strength leads to an increase in the radiation emission
amplitude, i.e.\begin{equation}
\frac{\partial E}{\partial h}>0.\label{eq:Radiation emission}\end{equation}
For a fixed driving strength, an increase in damping strength leads
to a decrease in the radiation emission amplitude, i.e.\[
\frac{\partial E}{\partial\gamma}<0.\]

It is also interesting to note that the velocity of radiation waves
is the same for all damping and driving strengths, provided that the
radiation wave is sufficiently separated from the soliton core. These
results of the dependence of soliton attractors on the damping and
driving strengths are summarized in Table~\ref{tab:Table}.

\begin{table}[t]
\begin{centering}
\begin{tabular}{|c|c|c|}
\hline 
\textbf{\large Characteristic} & \textbf{\large Increase: Damping} & \textbf{\large Increase: Driving}\tabularnewline
\hline 
Temporal Period $\left(T\right)$ & Decrease & Decrease\tabularnewline
\hline 
Magnitude of temporal oscillation $(m)$ & Decrease & Increase\tabularnewline
\hline 
Radiation emission $(E)$ & Decrease & Increase\tabularnewline
\hline 
Velocity of radiation waves & No effect & No effect\tabularnewline
\hline
\end{tabular}
\par\end{centering}

\caption{\emph{Dependence of soliton attractors on damping and driving strengths}\label{tab:Table}}
\end{table}

Another aspect of radiation waves that is influenced by the driving
strength is the position where the radiation waves are formed. Radiation
waves that form in the tail of the soliton can be easily identified.
Moreover, if radiation waves form in the tail and move away from the
quiescent soliton, it is reasonable to expect little interaction with
the soliton part of the attractor. Conversely if the radiation waves
form inside the soliton, it complicates the analysis of both the soliton
part and the radiation part of the soliton attractor, as well as the
interaction between the two components.

\begin{figure}[t]
\begin{centering}
\includegraphics[scale=0.3]{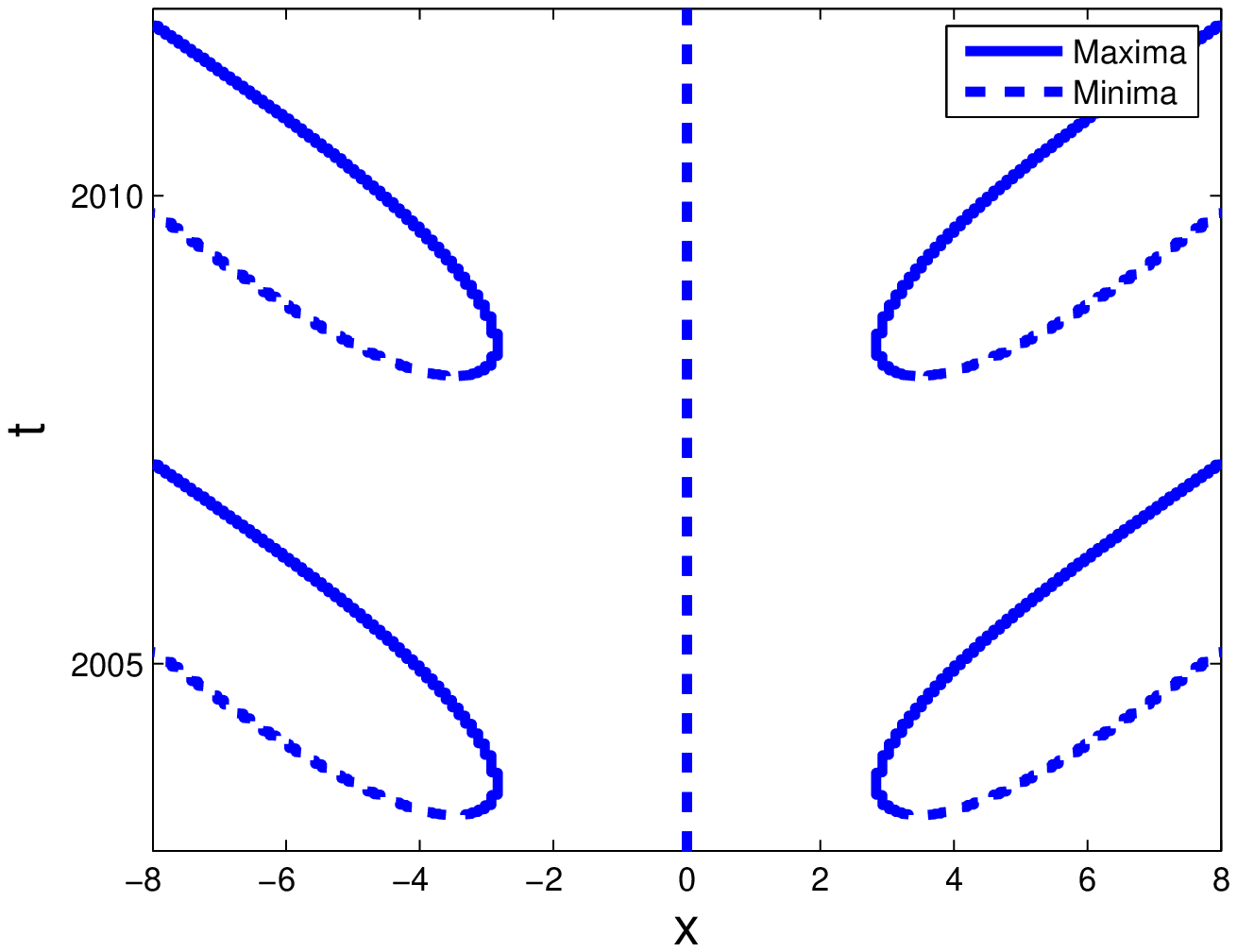}\includegraphics[scale=0.3]{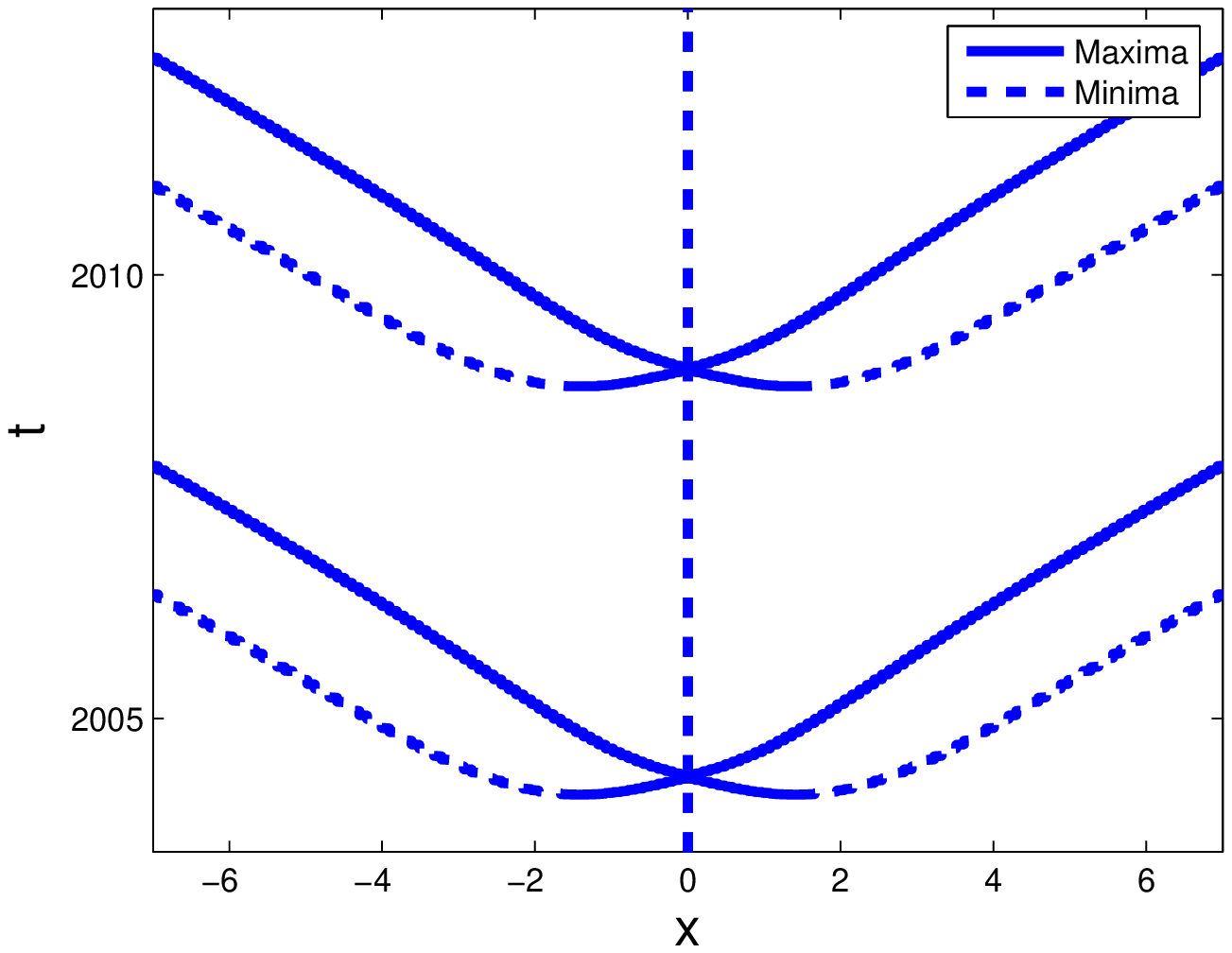}\includegraphics[scale=0.3]{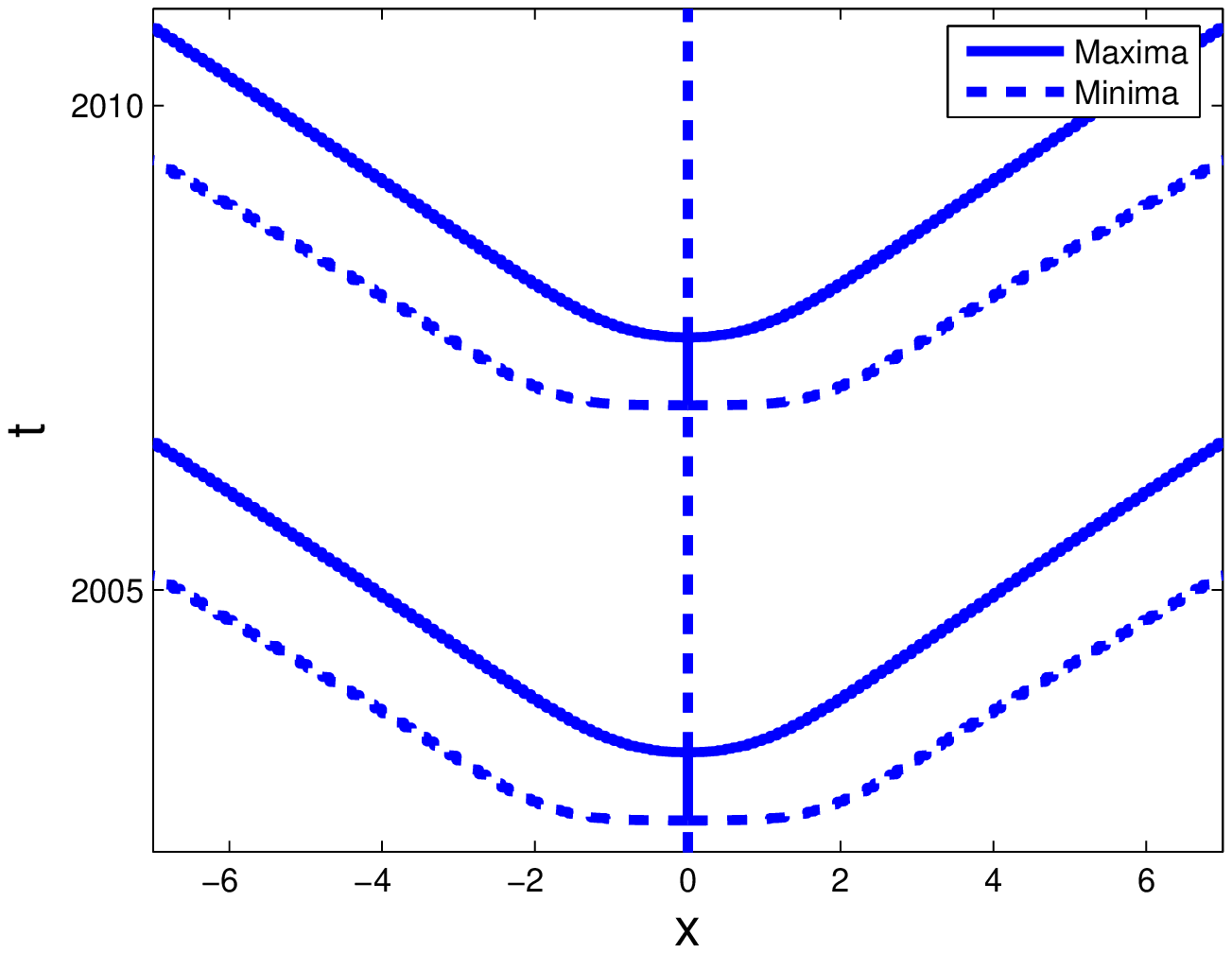}
\par\end{centering}

\begin{centering}
\emph{(a)~~~~~~~~~~~~~~~~~~~~~~~~~~~~~(b)~~~~~~~~~~~~~~~~~~~~~~~~~~~~~(c)}
\par\end{centering}

\caption{\emph{\small The three different ways that radiation is emitted. $h=0.39$
in (a), $h=0.46$}{\small{} }\emph{\small in (b), and $h=0.53$ in
(c).}{\small \label{fig:Extrema}}}
\end{figure}

Numerical results show that, for smaller driving strengths, radiation
waves are formed on the tail of the soliton. However, for larger driving
strengths, radiation waves are formed inside the soliton core. To
illustrate this, we consider the imaginary part of the soliton attractors
that arises for damping strength $\gamma=0.3$. We use the spatial
profile to show the formation and behaviour of the radiation waves.
The spatial profile consists of all pairs $\left(x,t\right)$ that
are local extremes with respect to $x,$ i.e. pairs that satisfy\[
\frac{\partial\psi^{(m)}}{\partial x}=0.\]
Solid lines represent local minima, and dotted lines denote local
maxima. In Figure~\ref{fig:Extrema} we show the spatial profiles
of attractors that arise for a damping strength of $\gamma=0.3$ and
for different driving strengths. In Figure~\ref{fig:Extrema}~(a)
we see that the radiation waves form at $x\approx\pm3.5$. In this
case the radiation waves simply move away from the soliton. However,
as driving strength increases, the distance between the point of formation
and the soliton decreases. This is shown in Figure~\ref{fig:Extrema}~(b)
where the driving strength is $h=0.46$. Here we see that the radiation
waves form at $x\approx\pm1.5$. Moreover, we see that the local maxima
move through the soliton. When the driving strength is increased further,
the radiation waves form at the origin, the same position as the centre
of the soliton. This is shown in Figure~\ref{fig:Extrema}~(c),
where the spatial profile is shown for $h=0.53$. Here we see that
both radiation waves are formed at the origin.

\section{The direct scattering study of soliton attractors\label{sec:Direct scattering study}}

The complex formation of radiation waves within the soliton core,
associated with larger driving strengths, motivates the use of the direct
scattering study. The inverse scattering transform method is an analytical
method for solving the focusing NLS equation. The method maps the
initial condition onto an auxiliary space where the transformed set,
known as the scattering data, evolves in a trivial manner. The evolved
scattering data can then be used to reconstruct the solution. An important
aspect about the mapping of the potential onto the scattering data
is that it divides the potential into a discrete set and a continuous
set. The discrete set describes the soliton part of the solution,
while the continuous set describes the radiation part of the solution.
For soliton attractors of the PDNLS equation the scattering data can
be used as a filter to separate the soliton part from the radiation
part. 

The discrete eigenvalues of the associated Zakharov-Shabat (ZS) eigenvalue
problem form part of the discrete scattering data. The ZS eigenvalue
problem is defined as\begin{equation}
\begin{array}{ccc}
v_{1x} & = & -i\zeta v_{1}+\psi v_{2}\\
v_{2x} & = & -\psi^{*}v_{1}+i\zeta v_{2}\end{array}.\label{eq:ZS}\end{equation}
Here $\psi$ is a solution of the PDNLS equation at a fixed point
in time and $\zeta$ is an eigenvalue. The discrete eigenvalues of
the ZS eigenvalue problem \eqref{eq:ZS} are associated with eigenfunctions
$v_{1,2}$ that decay to zero when $\left|x\right|\rightarrow\infty$.
We refer to the discrete eigenvalues as the ZS eigenvalues. The set
of ZS eigenvalues are referred to as the soliton content, due to its
relationship with solitons. In the literature these eigenvalues are
also sometimes referred to as nonlinear modes. Their relationship
with solitons is briefly discussed in the appendix.

We calculated the soliton content numerically by using the optimal
Floquet exponent of Hill's method, described in \cite{key-35}. The
results reveal that there are four different types of soliton attractors,
each with a unique behaviour pattern with respect to the soliton content.
After briefly illustrating the different types of attractors in Sections~3.1
-- 3.4, we discuss the general structure of soliton attractors in
Section~3.5.

\subsection{\noindent \textit{\emph{Type I attractor}}}

\begin{figure}[t]
\begin{centering}
\includegraphics[scale=0.4]{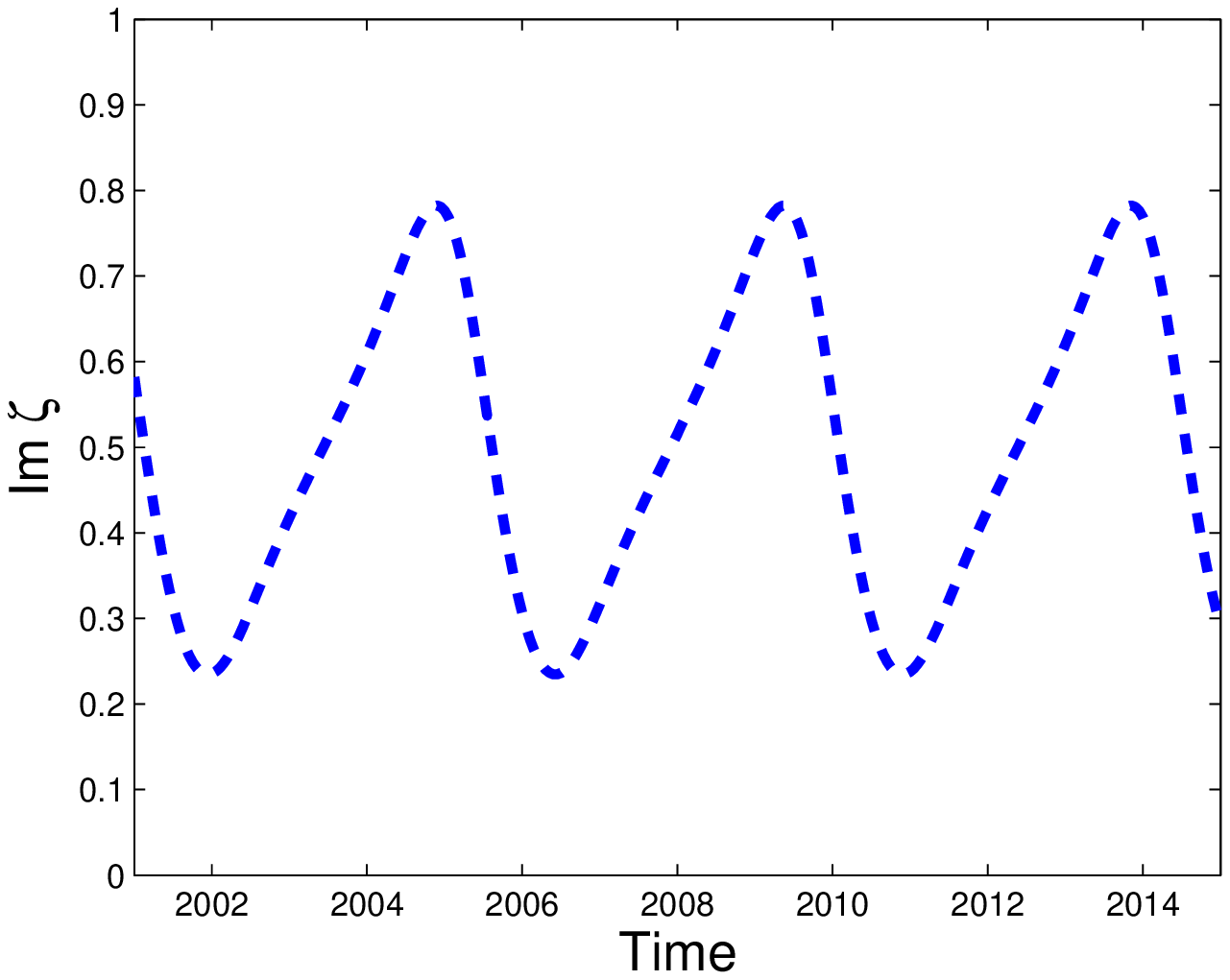} ~~~~~\includegraphics[scale=0.4]{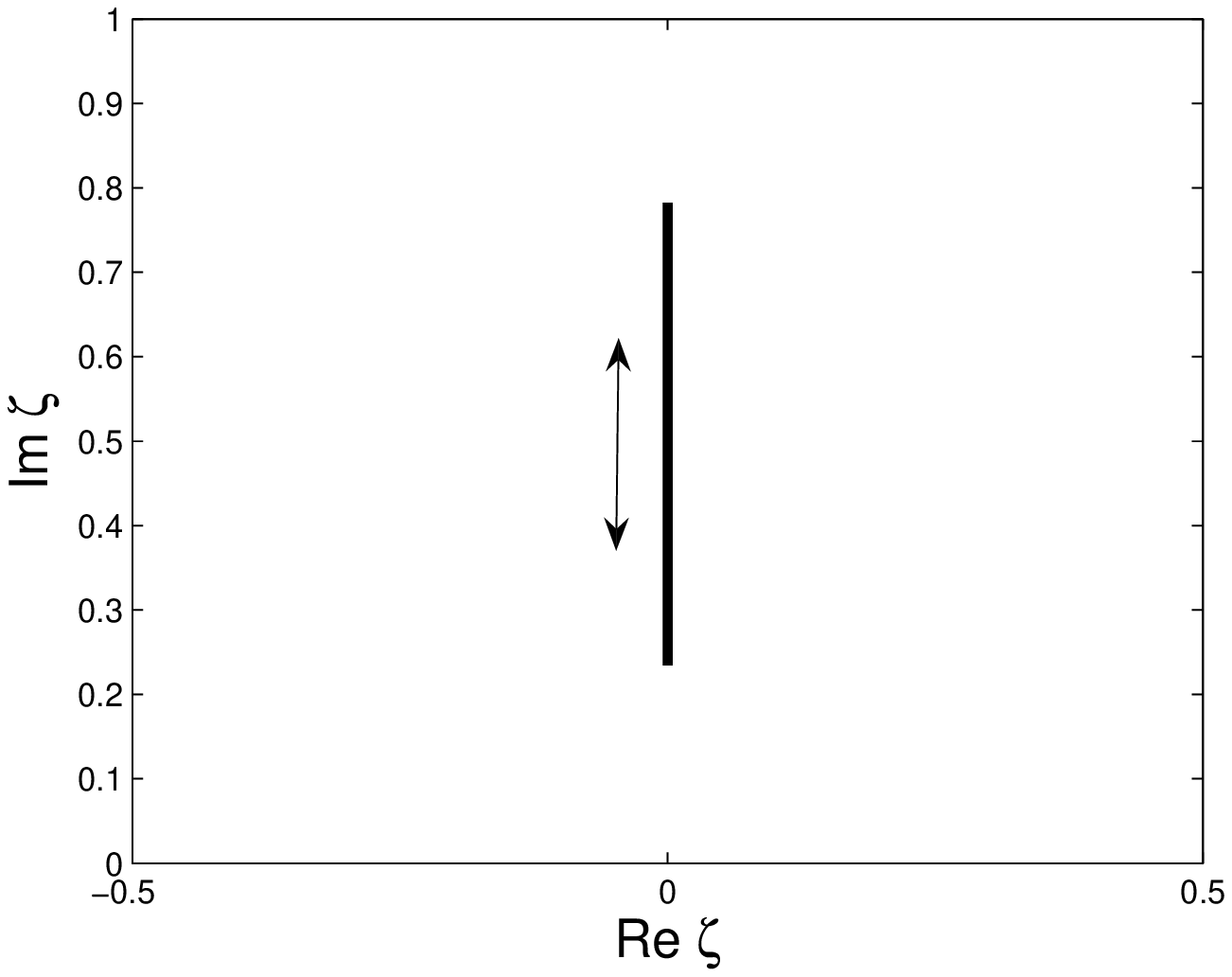}
\par\end{centering}

\begin{centering}
\emph{(a)~~~~~~~~~~~~~~~~~~~~~~~~~~~~~~~~~~~~~~~~~~~~~~~~~(b)}
\par\end{centering}

\begin{centering}
\includegraphics[scale=0.4]{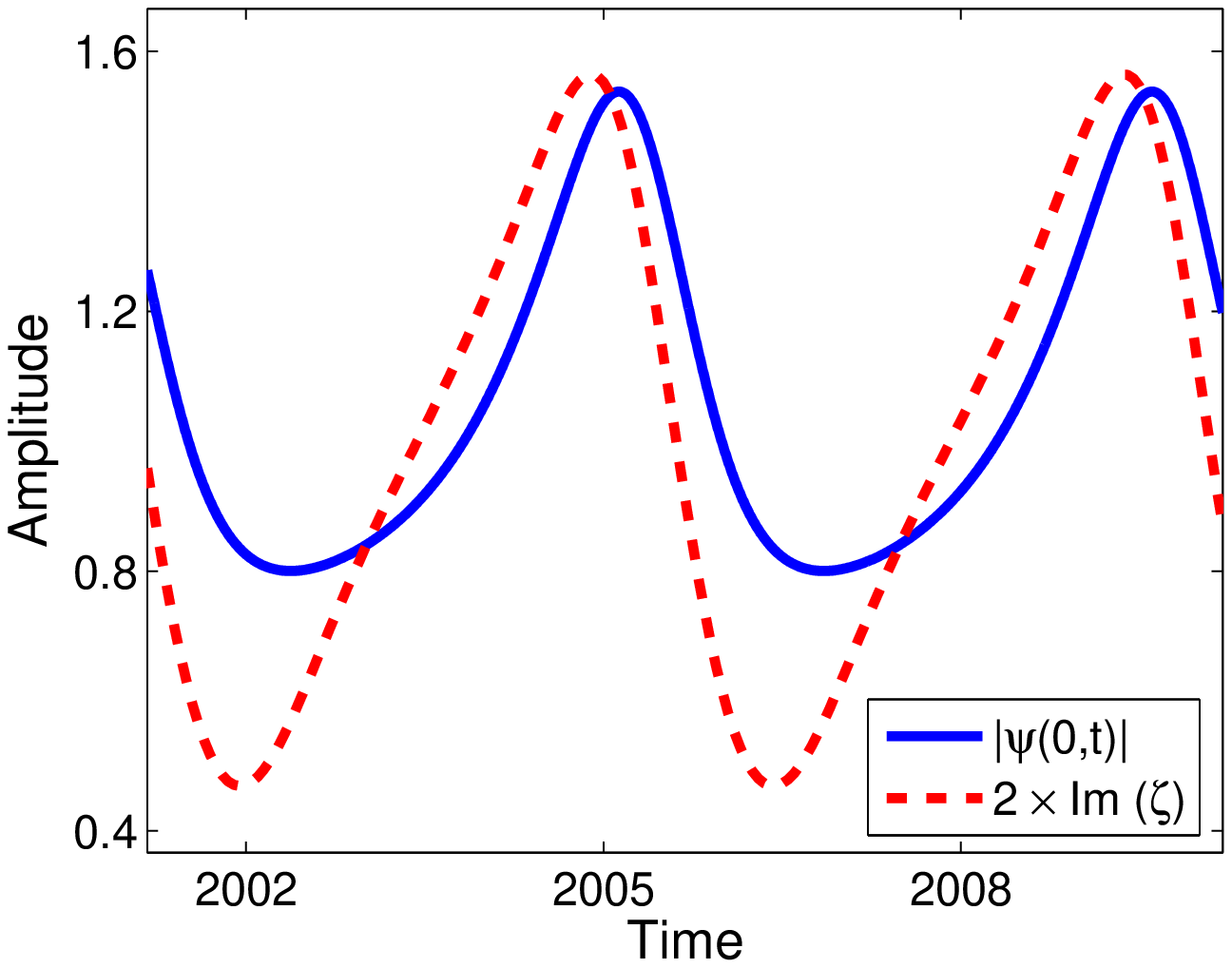}
\par\end{centering}

\begin{centering}
\emph{(c)}
\par\end{centering}

\caption{\label{fig:TI}\emph{Type I attractor, corresponding to $\gamma=0.3$,
$h=0.51$. The time evolution of the imaginary part of the ZS eigenvalue
is shown in (a). In (b) we show the image of the ZS eigenvalue on
the complex plane. In (c) we show the amplitudes of the attractor
(solid line) and the reconstructed soliton (dotted line).}}
\end{figure}
\textit{\emph{The soliton content of Type~I attractors consists of
a single purely imaginary ZS eigenvalue, oscillating on the imaginary
axis. Figure~\ref{fig:TI} shows the behaviour of the soliton content
for the soliton attractor that arises for damping and driving strengths
of $\gamma=0.3$ and $h=0.51$ respectively. Figure~\ref{fig:TI}~(a)
shows the time evolution of the imaginary part of the ZS eigenvalue.
In Figure~\ref{fig:TI}~(b) we plot the range of the ZS eigenvalues
on the complex plane, showing the path of the ZS eigenvalue on the
imaginary axis. In Figure~\ref{fig:TI}~(c) we compare the amplitudes
of the Type~I attractor (solid line) and the reconstructed soliton
associated with the ZS eigenvalue (dotted line). We see that, at its
minimum, the amplitude of the reconstructed soliton is much smaller
than the soliton attractor itself. This shows that the soliton part
of the attractor is in a dispersive state that is associated with
the formation and emission of radiation waves.}}

\subsection{\noindent \textit{\emph{Type II attractor}}}

\begin{figure}[t]
\begin{centering}
\includegraphics[scale=0.4]{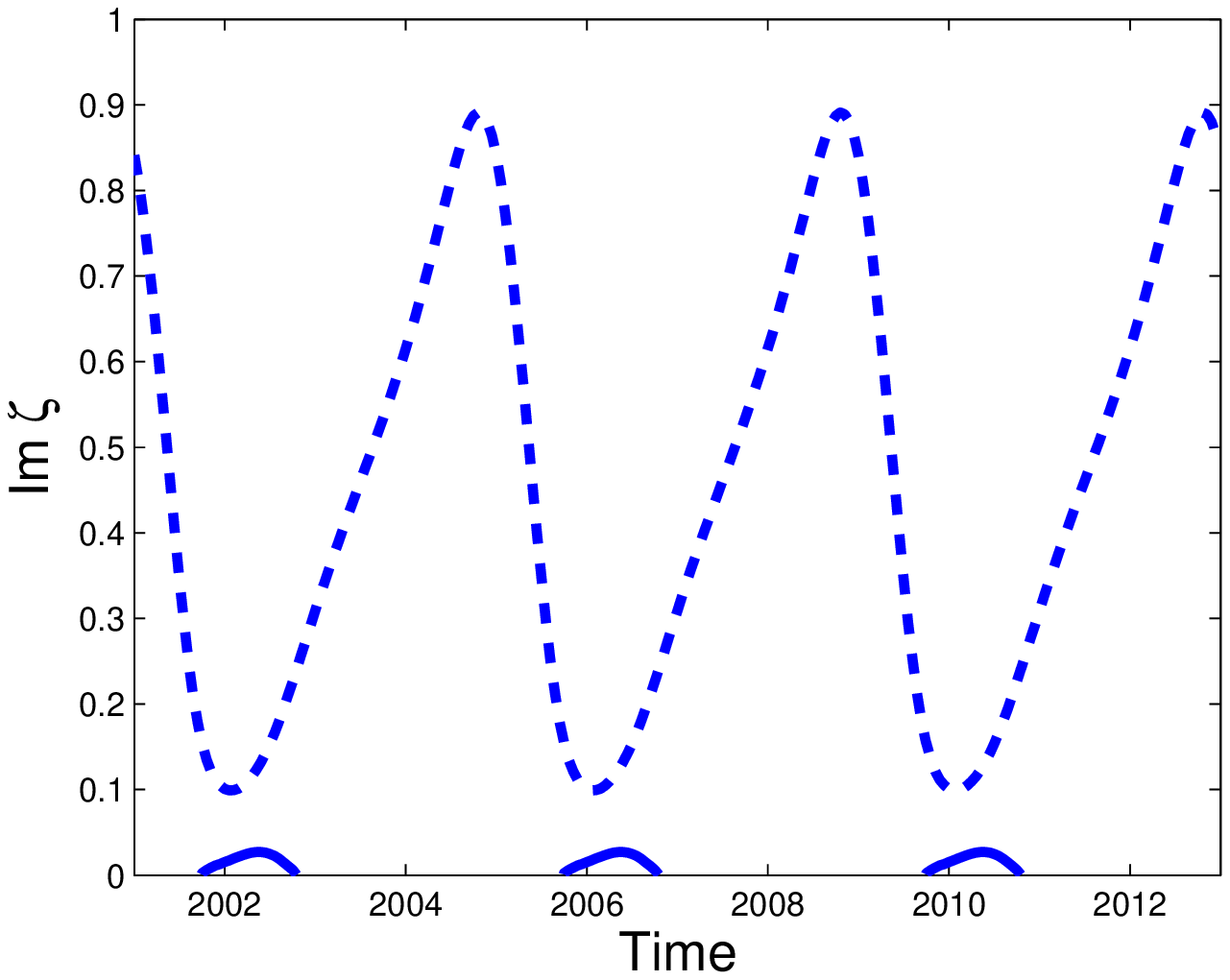}~~~~~\includegraphics[scale=0.4]{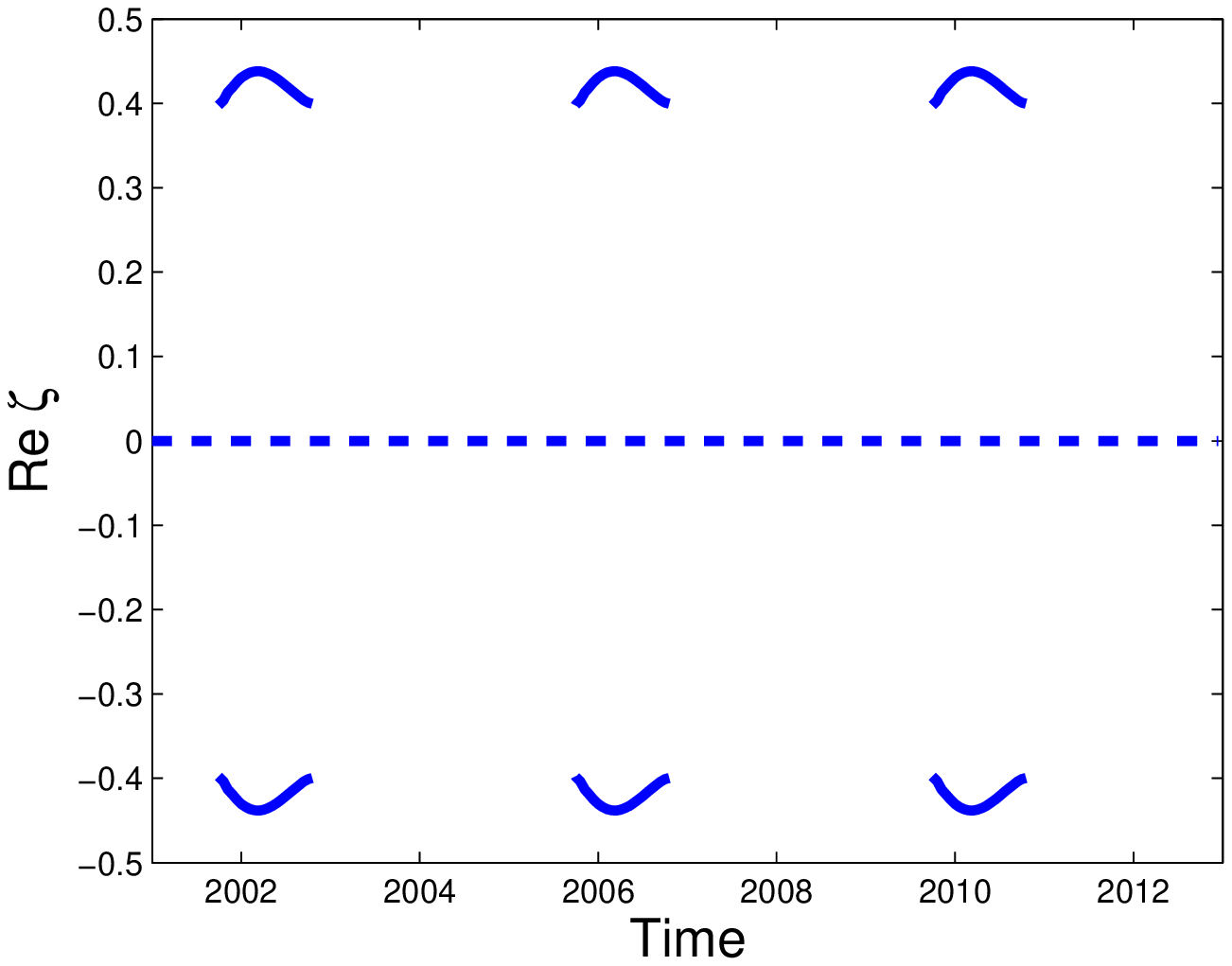} 
\par\end{centering}

\begin{centering}
\emph{(a)~~~~~~~~~~~~~~~~~~~~~~~~~~~~~~~~~~~~~~~~~~~~~~~~~~(b)}
\par\end{centering}

\begin{centering}
\includegraphics[scale=0.4]{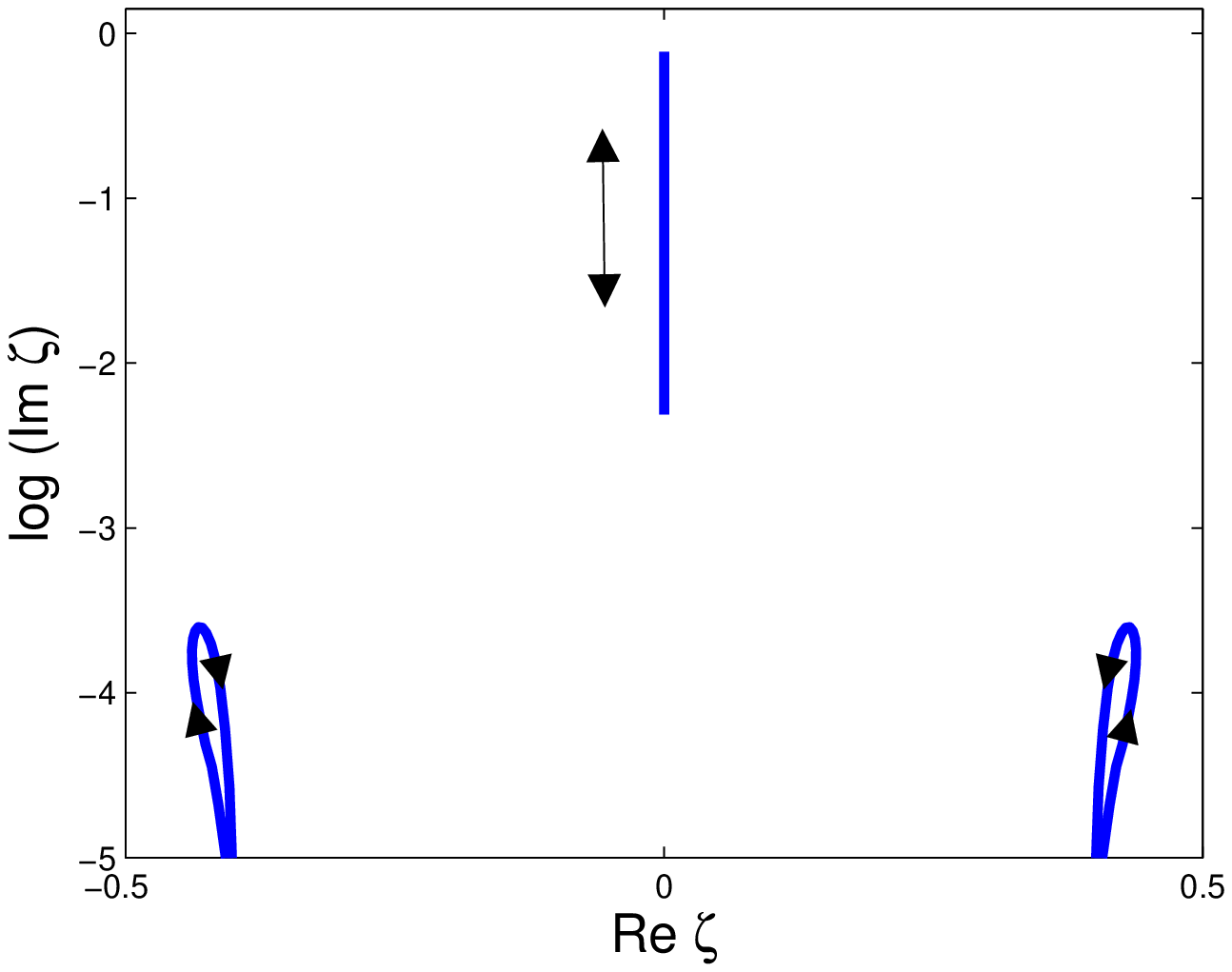}~~~~~\includegraphics[scale=0.4]{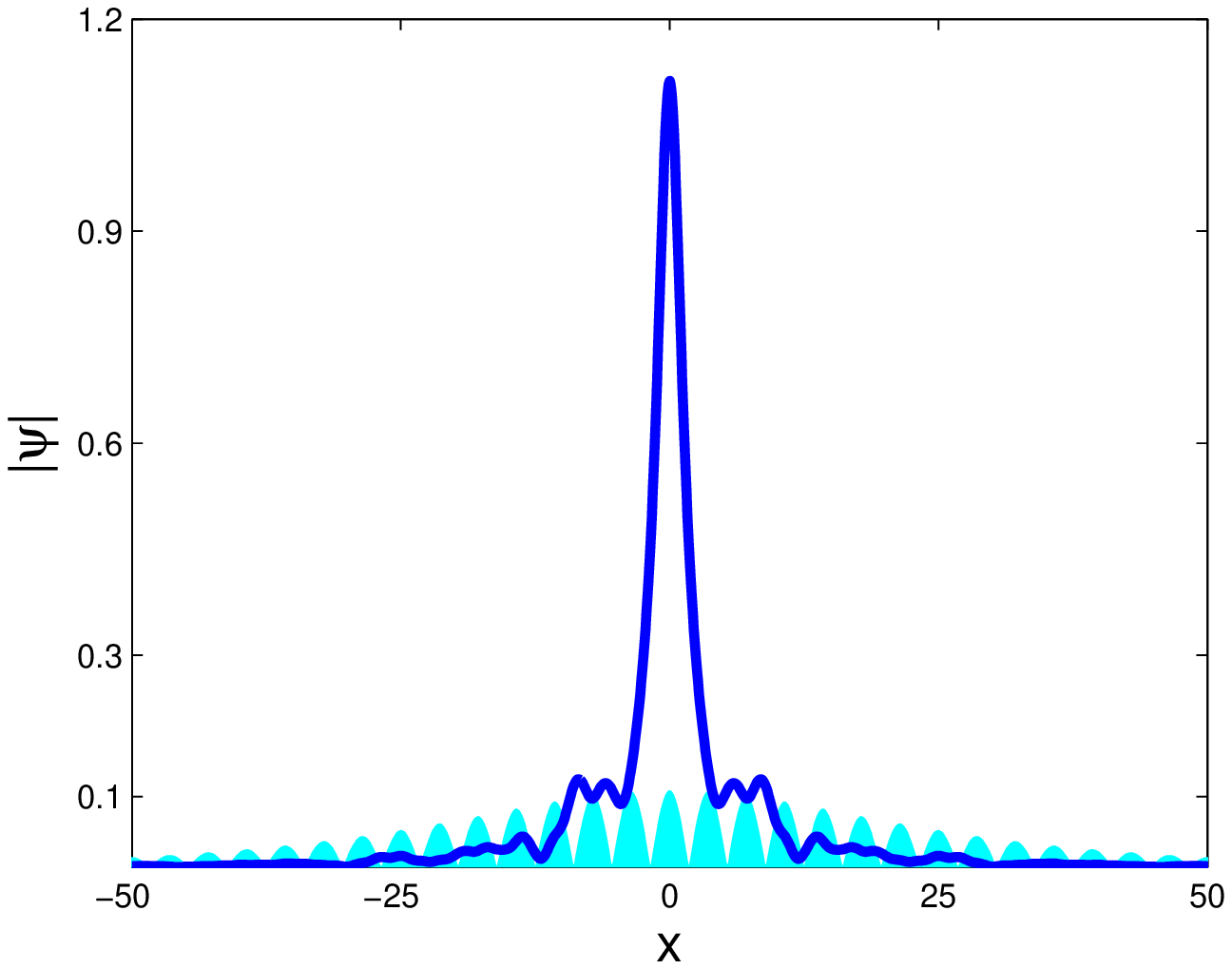}
\par\end{centering}

\begin{centering}
\emph{(c)~~~~~~~~~~~~~~~~~~~~~~~~~~~~~~~~~~~~~~~~~~~~~~~~~~(d)}
\par\end{centering}

\caption{\emph{\label{fig:TII}Type II attractor, corresponding to $\gamma=0.3$,
$h=0.65$. The time evolution of the imaginary parts and the real
parts of the ZS eigenvalues is shown in (a) and (b) respectively.
In (c) the image of the ZS eigenvalues on the complex plane is shown.
In (d) the reconstructed soliton associated with the side eigenvalues
(shaded area) and the solution is shown at $t=2002.35$.}}
\end{figure}

\noindent \textit{\emph{The soliton content of Type~II attractors
consists of one oscillating purely imaginary ZS eigenvalue and a pair
of two complex ZS eigenvalues that appear and disappear periodically.
The pair of ZS eigenvalues is symmetric about the imaginary axis.
We refer to them as }}\textit{side eigenvalues}\textit{\emph{. Figure~\ref{fig:TII}
shows the soliton content of the Type II attractor that arises for
damping and driving strengths $\gamma=0.3$ and $h$~$=$~$0.65$
respectively. Figure~\ref{fig:TII}~(a) shows the time evolution
of the imaginary part of the soliton content. We use dotted lines
to represent purely imaginary ZS eigenvalues and solid lines to represent
ZS eigenvalues with non-zero real parts. We see that the side eigenvalues
appear and disappear in the region where the purely imaginary eigenvalue
approaches a local minimum, corresponding to the time when radiation
waves are emitted. Therefore side eigenvalues are associated with
a state of large radiation emissions. In Figure~\ref{fig:TII}~(b)
the real part of the discrete spectrum is shown, illustrating the
symmetry of the side eigenvalues. Here we see that the side eigenvalues
are not formed at the origin, but on the real line at $\zeta\approx\pm0.4$.
In Figure~\ref{fig:TII}~(c) we show the range of the ZS eigenvalues
in the complex plane. We used a logarithmic scale of the imaginary
part of the eigenvalue to emphasize the behaviour of the side eigenvalues.}}
The relationship between side eigenvalues and radiation is illustrated
in \textit{\emph{Figure~\ref{fig:TII}~(d). Here the solid line
shows the solution $\psi$ at $t=2002.35$. On this graph we use the
shaded area to superimpose the solution that is reconstructed from
the side eigenvalues (see }}appendix for more details). Note that
that the reconstructed solution is well correlated with the radiation
tail of the solution.

\subsection{\noindent \textit{\emph{Type III attractor}}}

\begin{figure}[t]
\begin{centering}
\includegraphics[scale=0.4]{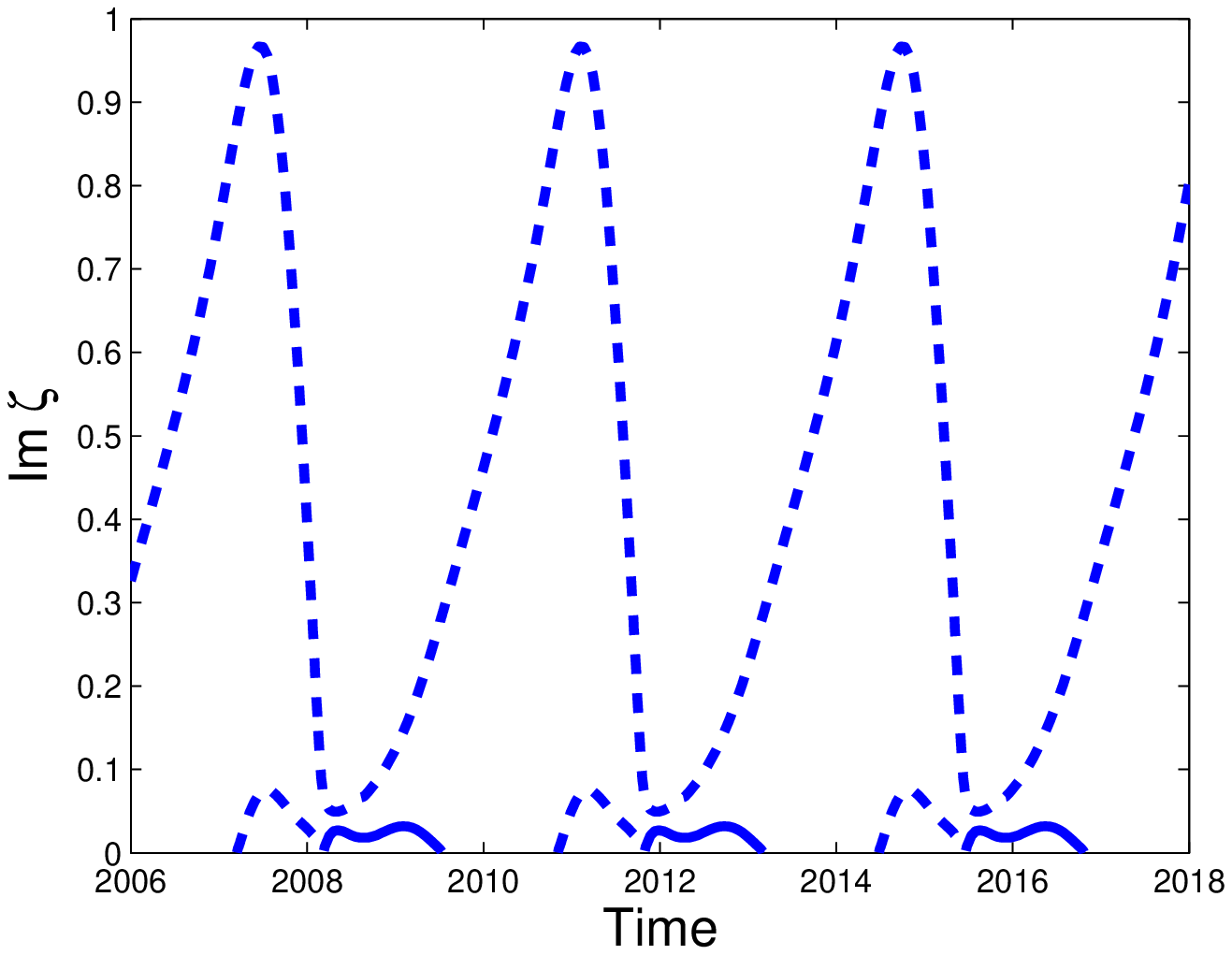}~~~~~\includegraphics[scale=0.4]{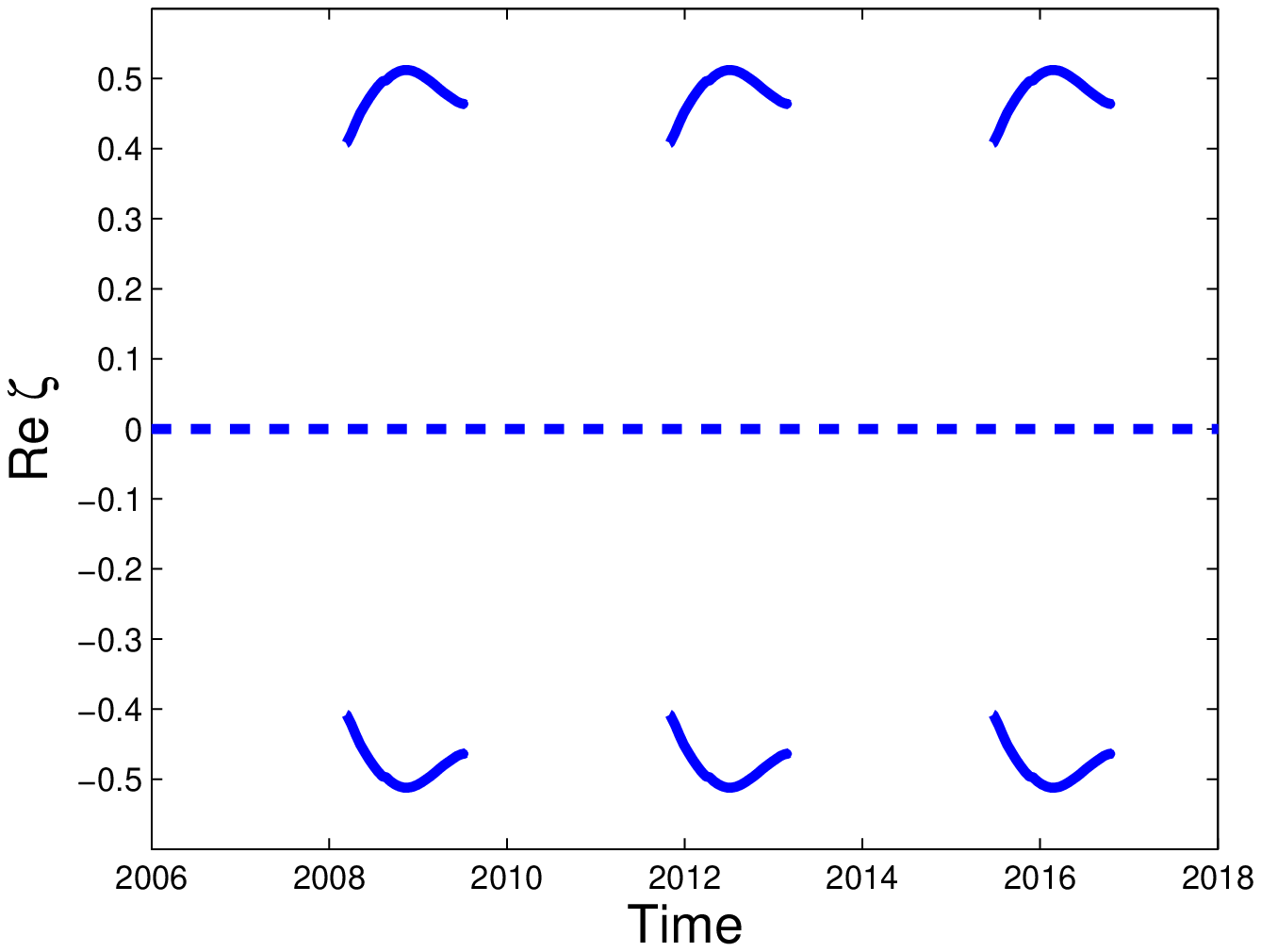}
\par\end{centering}

\begin{centering}
\emph{(a)~~~~~~~~~~~~~~~~~~~~~~~~~~~~~~~~~~~~~~~~~~~~~~~~~(b)}
\par\end{centering}

\begin{centering}
\includegraphics[scale=0.4]{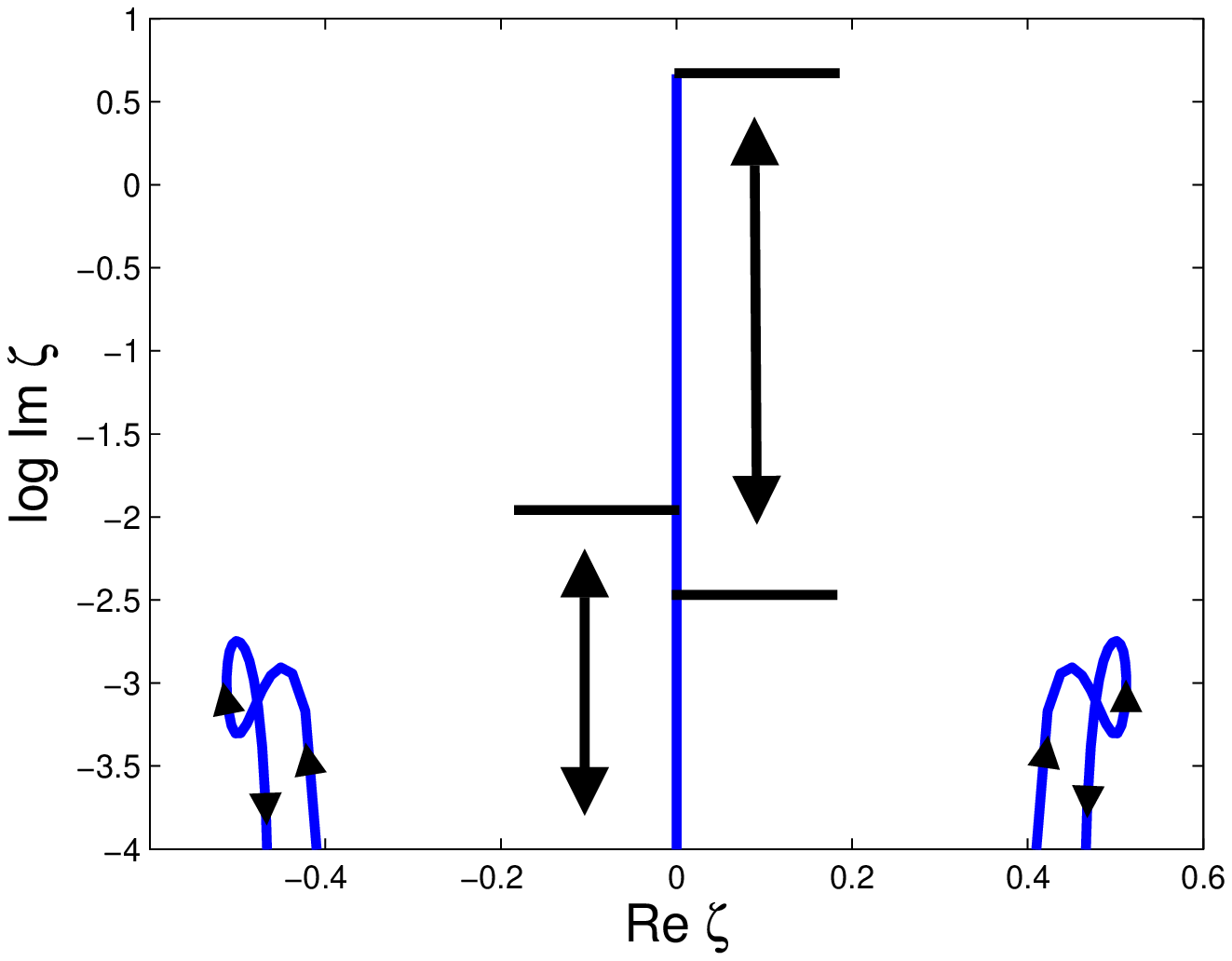}~~~~~\includegraphics[scale=0.4]{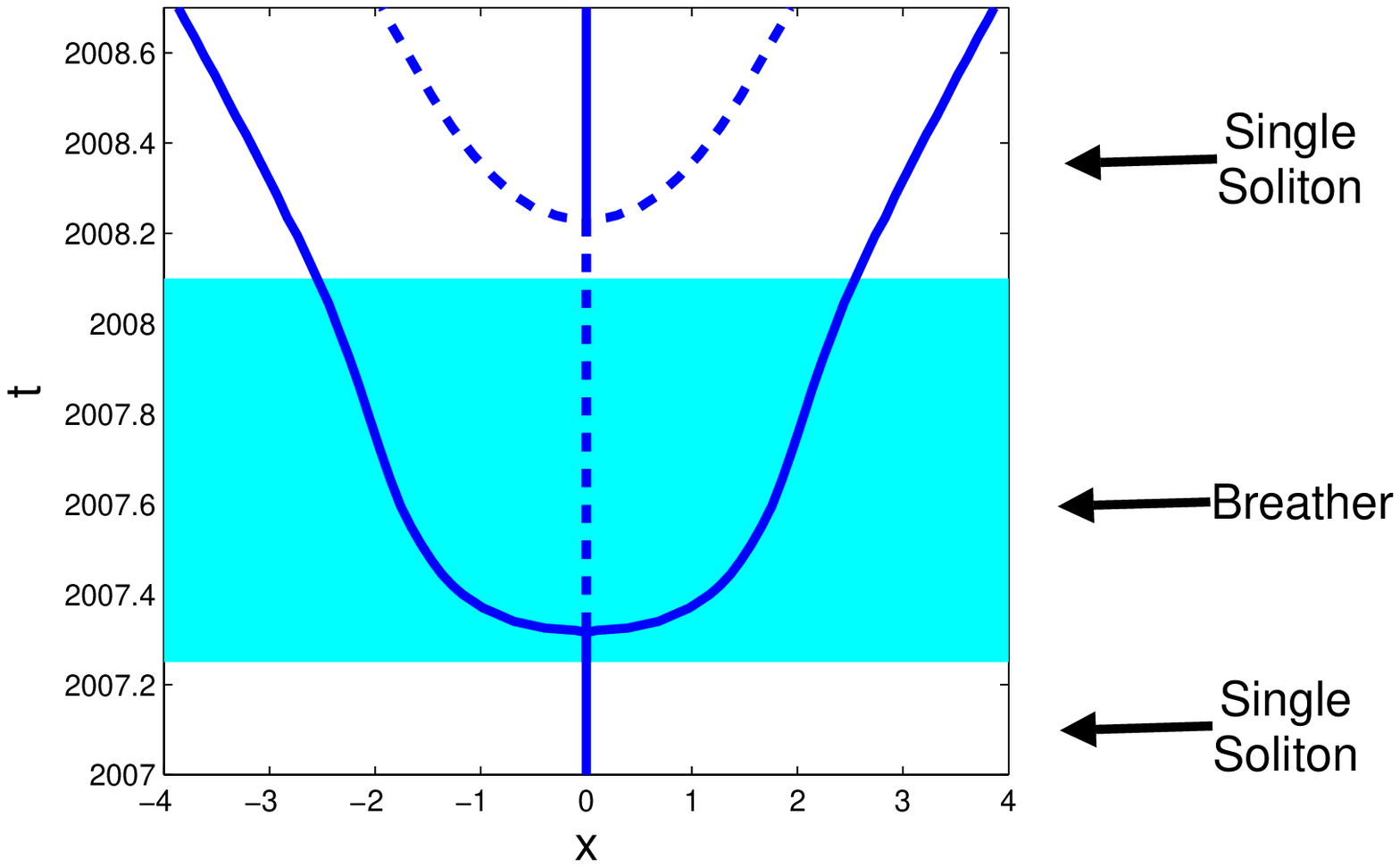}
\par\end{centering}

\emph{~~~~~~~~~~~~~~~~~~~~~~~~~~(c)~~~~~~~~~~~~~~~~~~~~~~~~~~~~~~~~~~~~~~~~~~~~~~(d)}

\caption{\label{fig:TIII}\emph{Type III attractor, corresponding to $\gamma=0.3$,
$h=0.77$. The time evolution of the imaginary parts and the real
parts of the ZS eigenvalues are shown in (a) and (b) respectively.
In (c) the image of the ZS eigenvalues on the complex plane is shown.
In (d) the spatial profile of the imaginary part of the solution is
shown.}}
\end{figure}

\textit{\emph{The soliton content of Type~III attractors consists
of }}\textit{1)}\textit{\emph{ an oscillating purely imaginary ZS
eigenvalue }}\textit{2)}\textit{\emph{ side eigenvalues and }}\textit{3)}\textit{\emph{
a second purely imaginary ZS eigenvalue that appears and disappears
periodically. In Figure~\ref{fig:TIII} the soliton content is shown
for the attractor that arises for damping and driving strengths $\gamma=0.3$
and $h=0.77$ respectively. Figure~\ref{fig:TIII}~(a) shows the
time evolution of the purely imaginary part of the ZS eigenvalues.
Note that the second purely imaginary eigenvalue, marked with the
dotted line, appears when the original purely imaginary ZS eigenvalue
approaches its maximum. In Figure~\ref{fig:TIII}~(b) the time evolution
of the real part of the discrete eigenvalues are shown. Here we see
the side eigenvalues appearing and disappearing periodically. In Figure~\ref{fig:TIII}~(c)
we show the range of the ZS eigenvalues. Since the paths of the purely
imaginary ZS eigenvalues overlap, we show the range of the larger
(and permanent) purely imaginary ZS eigenvalue with the arrow on the
right hand side. We use the arrow on the left to indicate the path
of the smaller (and temporary) purely imaginary ZS eigenvalue. Once
again we plotted the imaginary axis on the logarithmic scale to emphasize
the behaviour of the smaller ZS eigenvalues.}}

\textit{\emph{The formation of a second purely imaginary ZS eigenvalue
is associated with the formation of breather-like lateral waves. This
is illustrated in Figure~\ref{fig:TIII}~(d) where we show the spatial
profile of the imaginary part of the attractor. We see the formation
of two waves that move away from the soliton at $t\approx2007.25$.
During the lifetime of the breather (shaded area) their speed decreases.
This shows that they are coupled with the soliton to form breather-like
structures, consisting of a soliton and two lateral waves (see appendix).
Once the second purely imaginary ZS eigenvalue disappears, their speed
increases until they behave like regular radiation waves.}}

\subsection{\noindent \textit{\emph{Type IV attractor}}}

\begin{figure}[tb]
\begin{centering}
\includegraphics[scale=0.4]{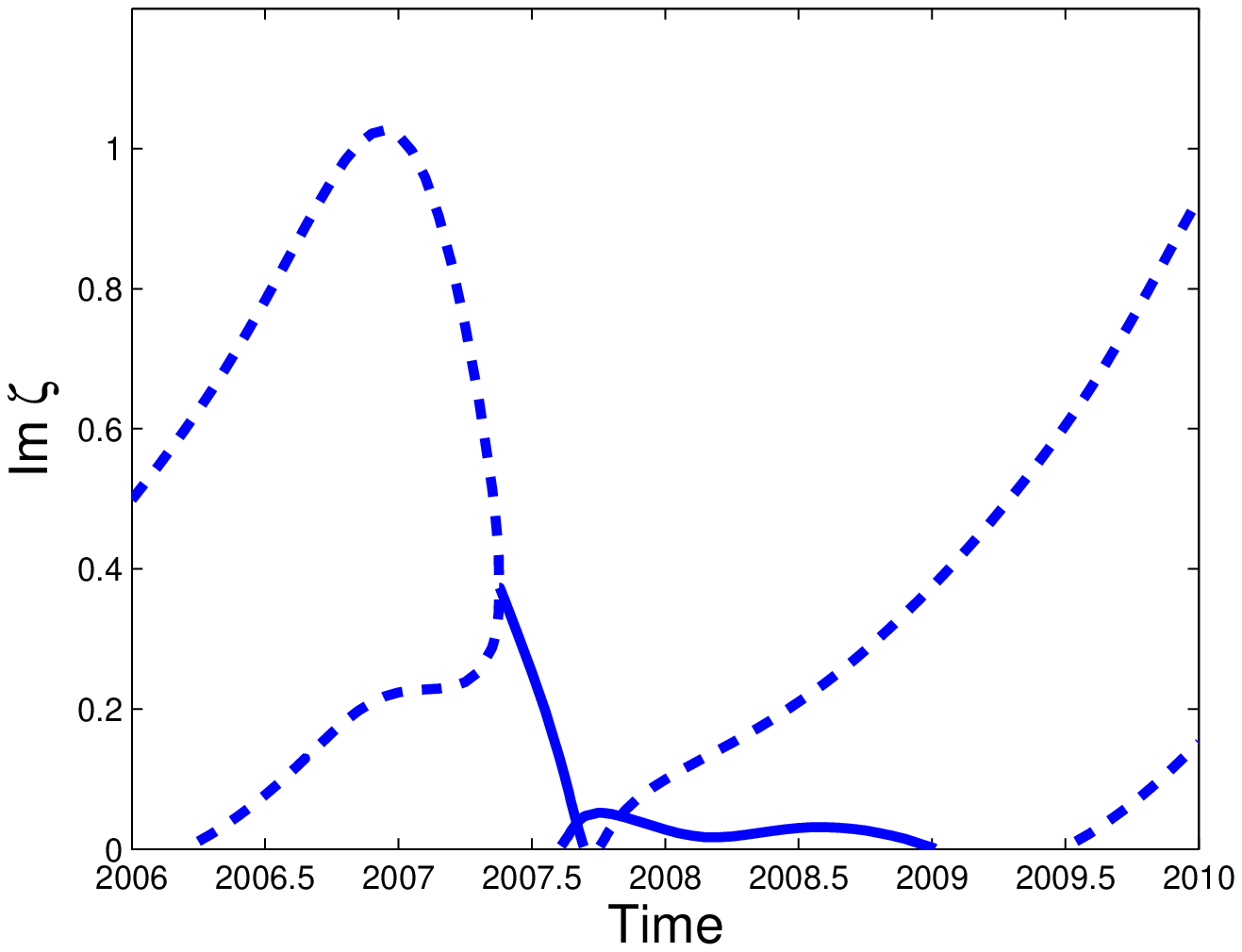}~~~~~\includegraphics[scale=0.4]{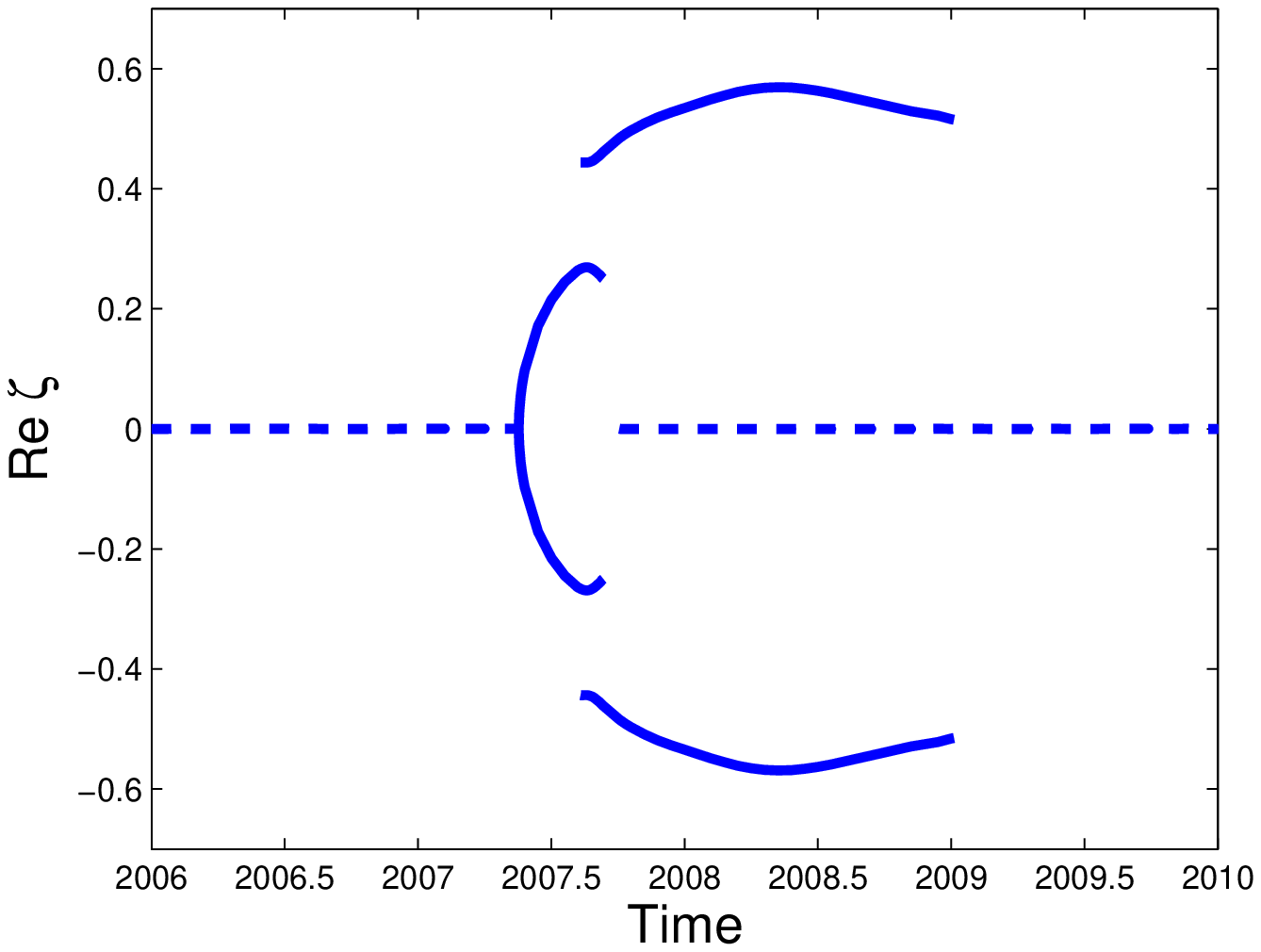}
\par\end{centering}

\begin{centering}
\emph{(a)~~~~~~~~~~~~~~~~~~~~~~~~~~~~~~~~~~~~~~~~~~~~~~~~~(b)}
\par\end{centering}

\begin{centering}
\includegraphics[scale=0.4]{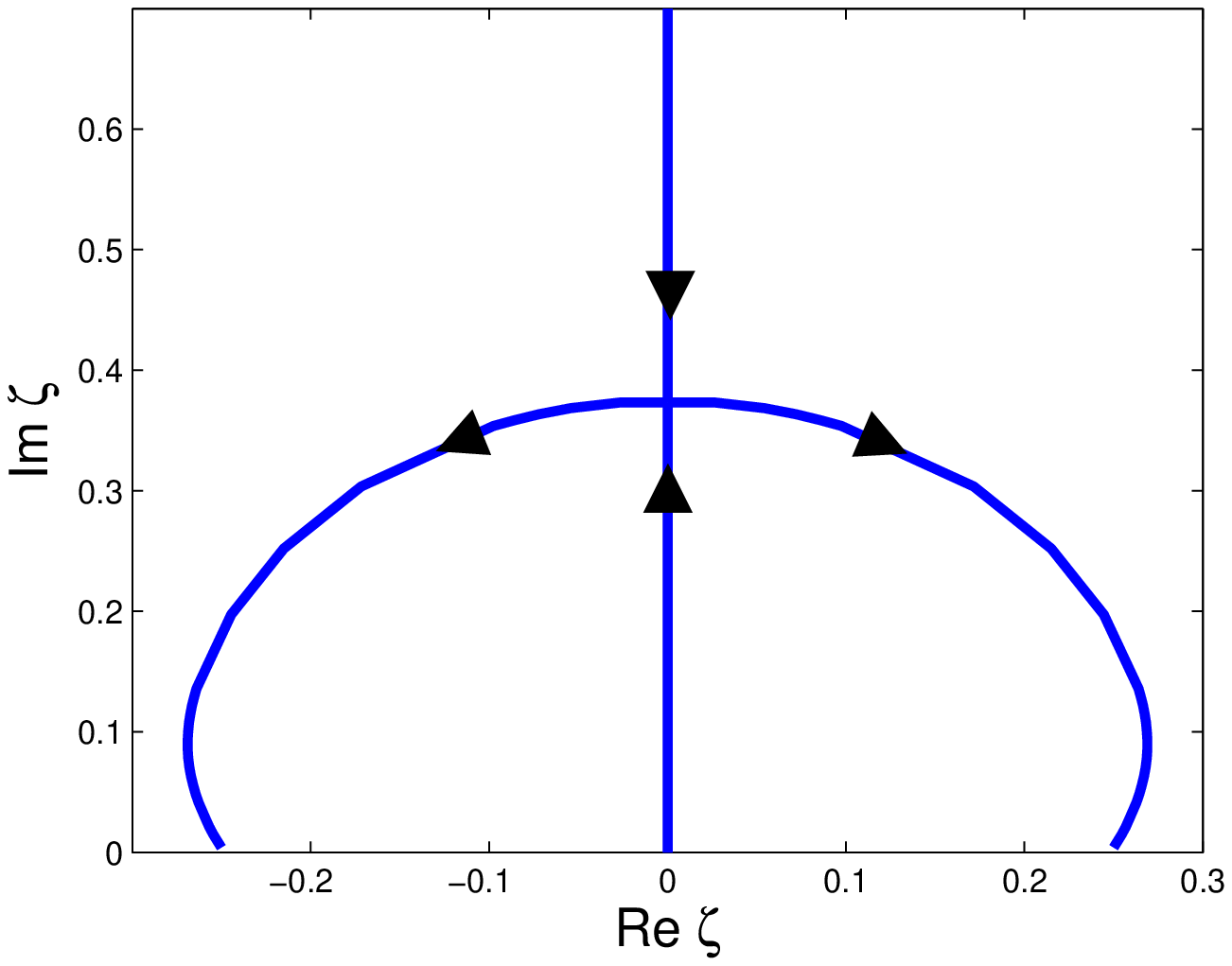}~~~~~\includegraphics[scale=0.4]{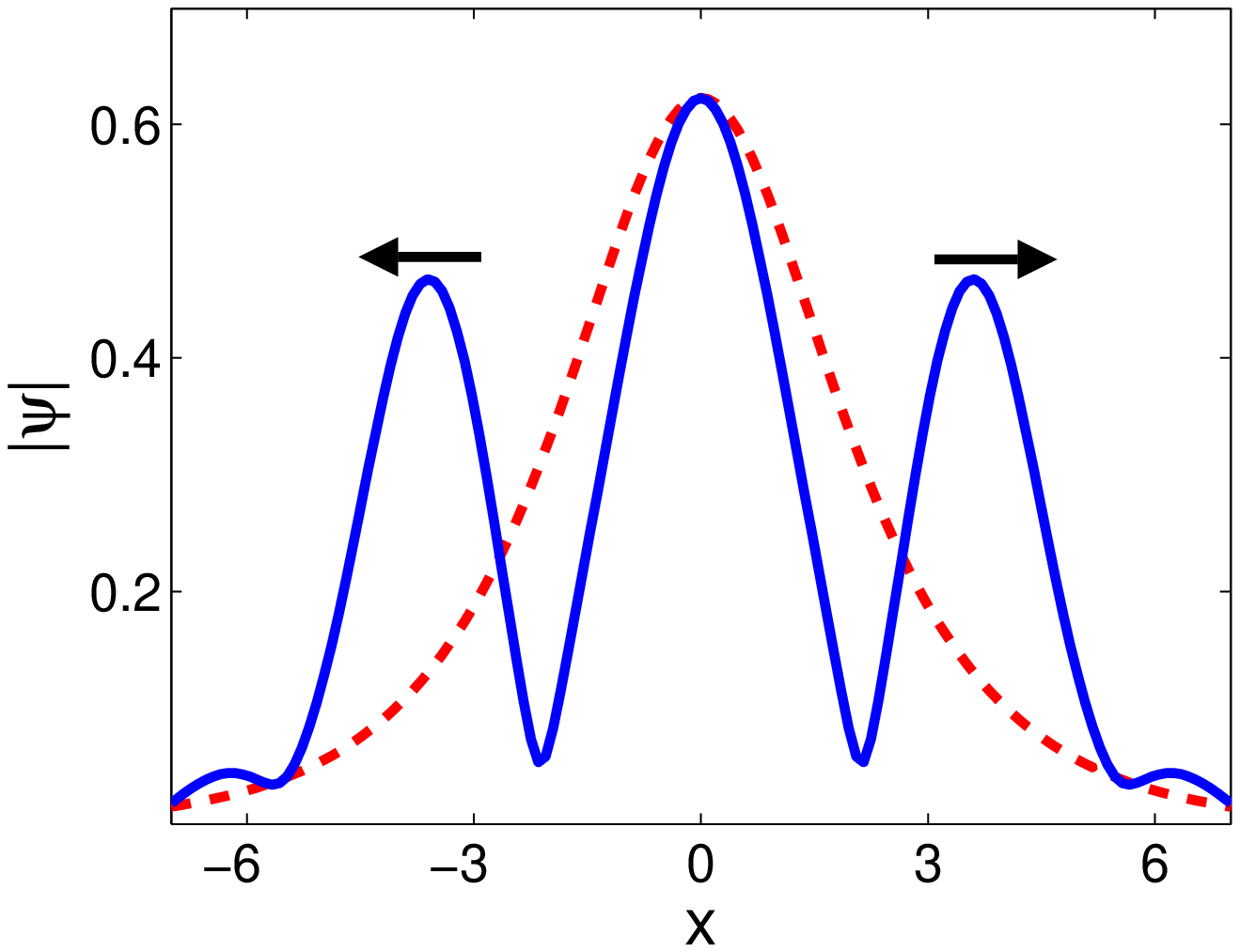}
\par\end{centering}

\begin{centering}
\emph{(c)~~~~~~~~~~~~~~~~~~~~~~~~~~~~~~~~~~~~~~~~~~~~~~~~~(d)}
\par\end{centering}

\caption{\label{fig:TIV}\emph{Type IV attractor, corresponding to $\gamma=0.3$,
$h=0.87$. The time evolution of the imaginary parts and the real
parts of the ZS eigenvalues are shown in (a) and (b) respectively.
In (c) the image of the splitting eigenvalues on the complex plane
is shown. In (d) the solution is shown at $t=2008.15$ (solid) and a reconstructed soliton with same amplitude (dotted).}}
\end{figure}

\textit{\emph{The soliton content of Type~IV attractors differ from
that of Type~III attractors in the behaviour of the purely imaginary
ZS eigenvalues. In this case, the second purely imaginary ZS eigenvalue
does not disappear into the origin. Instead, the two imaginary ZS
eigenvalues collide. After the collision, the two eigenvalues move
into the complex plane, symmetric about the imaginary axis, and disappear
into the real line, in a similar way that the side eigenvalues disappear.
We refer to these eigenvalues as }}\textit{splitting eigenvalues.}\textit{\emph{
While this happens, a purely imaginary ZS eigenvalue forms at the
origin. In Figure~\ref{fig:TIV} we show the results for the Type~IV
attractor that arises for damping and driving strengths $\gamma=0.3$
and $h=0.87$ respectively. The time evolution of the imaginary parts
of the ZS eigenvalues is shown in Figure~\ref{fig:TIV}~(a). In
Figure~\ref{fig:TIV}~(b) we show the time evolution of the real
parts of the ZS eigenvalues. Here we see how the splitting eigenvalues
move symmetrically in the complex plane. }}In Figure~\ref{fig:TIV}~(c)
the behaviour of the splitting eigenvalues is shown for $2007<t<2008.$
Here we see how the two purely imaginary ZS eigenvalues collide on
the imaginary axis, and how they move into the complex plane before
disappearing into the real line. Note that the imaginary parts of
the split eigenvalues is initially large relative to their real parts.
As they approach the real line, the converse is true. This corresponds
to a transition from lateral waves structure to radiation waves (see
Appendix). Therefore the splitting eigenvalues are associated with
the breakup of the soliton. This is illustrated in Figure~\ref{fig:TIV}~(d)
where the solution is shown at $t=2008.15$, shortly after the disappearance
of the splitting eigenvalues. The dotted line shows the modulus of a single soliton with the
same amplitude as the central wave, constructed using  equation \eqref{eq:Single soliton}.
Note that the width of the latter is large in comparison to
the central wave. This, combined with the fact that the lateral waves are moving
away from the central wave, results in the absence of ZS eigenvalues.

\subsection{General structure of soliton attractor region}

\begin{table}[t]
\begin{centering}
\begin{tabular}{|c|c|c|c|c|}
\hline 
Attractor & Oscillating imaginary & Side & Second purely imaginary  & Splitting\tabularnewline
type & ZS eigenvalue & eigenvalues & ZS eigenvalue & eigenvalues\tabularnewline
\hline
\hline 
I & x &  &  & \tabularnewline
\hline 
II & x & x &  & \tabularnewline
\hline 
III & x & x & x & \tabularnewline
\hline 
IV &  & x & x & x\tabularnewline
\hline
\end{tabular}
\par\end{centering}

\caption{\emph{Characteristics of the soliton content of the different types
of attractors.}\label{tab:6.1}}
\end{table}

\begin{figure}[t]
\begin{centering}
\includegraphics[scale=0.6]{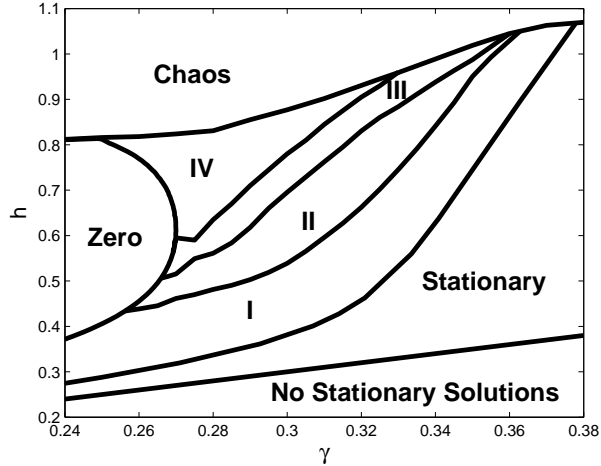}
\par\end{centering}

\caption{\label{fig:Map}\emph{Soliton structure chart of the }$\left(\gamma,h\right)$
\emph{plane}, \emph{showing the regions on which the Type I - IV attractors
exist.}}
\end{figure}

We calculated the soliton content throughout the oscillating soliton attractor
region. In Figure~\ref{fig:Map} the soliton structure chart shows
the distribution of different types of attractors. Here {}``\emph{I}''
corresponds to the Type~I attractor region etc. The intermediate
damping regime $0.27$~$\leq$~$\gamma$~$\leq$~$0.328$, where
all four types of soliton attractors arise, reveal a clear hierarchy,
namely that the Type I attractor region is bounded above by the Type
II attractor region, while the latter is bounded above by the Type
III attractor region. Finally the Type III attractor region is bounded
above by the Type IV attractor region. In Table~\ref{tab:6.1} we
summarize the characteristics of the soliton content associated with
each type of attractor.

The hierarchy of soliton attractors illustrates the effect of the
driving strength $h$ on soliton attractors. For driving strengths
near the Hopf bifurcation Type~I attractors arise. These attractors
consist of only a single purely imaginary ZS eigenvalue. An increase
in driving strength beyond the Type~I attractor region results in
the formation of side eigenvalues, associated with Type II attractors.
This shows that radiation emissions in this region are large relative
to those in the Type~I region. Indeed, throughout the Type~II, Type~III
and Type~IV regions, an increase in driving strength leads to an
increase in both the lifetime and maximum imaginary parts of the side
eigenvalues (see for example the imaginary parts of the side eigenvalues
in Figures~\ref{fig:TII} -- \ref{fig:TIV}). This indicates that
the radiation emissions increase when the driving strength is increased,
a result that is in good agreement with the radiation emission measurements
\eqref{eq:Radiation emission}.

When driving strength is increased beyond the Type~II attractor region,
a second purely imaginary ZS eigenvalue is formed in the soliton
content of the associated Type~III attractor. This shows that large
driving strengths affect the nature of the soliton itself. In particular,
during the lifetime of the second purely imaginary ZS eigenvalue,
the soliton couples with lateral waves to form a breather-like structure.
It is important to emphasize that the lateral waves form while the
purely imaginary ZS eigenvalue $\zeta_{im}$ reaches its maximum.
This shows that they are formed before the emission of radiation waves,
associated with $\zeta_{im}$ reaching its minimum. Indeed, as $\zeta_{im}$
decreases, the disappearance of the second purely imaginary ZS eigenvalue
shows that the breather-like structure gives way to a single soliton
structure. During this process the lateral waves become radiation
waves that propagate away from the soliton.

When the driving strength is increased beyond the Type~III region,
the purely imaginary ZS eigenvalues of the Type~IV attractor split
before disappearing into the real line. This is associated with a
breather-like structure that breaks up into radiation waves. The disappearance
of these ZS eigenvalues is balanced by the formation of new ZS eigenvalues
at the origin, associated with the formation of new solitons. In Section~\ref{sec:5}
we investigate the role of additional solitons in the formation of
spatio-temporal chaos.

In Figure~\ref{fig:Map} we see that the zero attractor region appears
in the small damping regime $\gamma<0.27$. It is interesting to note
that the zero attractor region is mostly bounded below by Type~I
and Type~II attractors. Therefore the formation of breather-like
structures doesn't seem to play a role in the formation of the zero
attractor region. In the following section we use the direct scattering
study to investigate the zero attractor region in more detail.

\section{The zero attractor region\label{sec:4}}

\subsection{The effect of period-doubling on radiation
emission \label{sub:Effect of period-doubling}}

The zero attractor region exists for damping strengths $\gamma<0.27$.
Oscillating soliton attractors {}``\emph{strike the balance between
energy fed by the driver and energy lost to dissipation}'' \cite{key-32}.
One therefore expects the zero attractor to arise as a result of more
energy lost through dissipation than energy fed by the driver. Since
radiation waves are dissipative, they play an important role in this
process.

\begin{figure}
\begin{centering}
\includegraphics[scale=0.4]{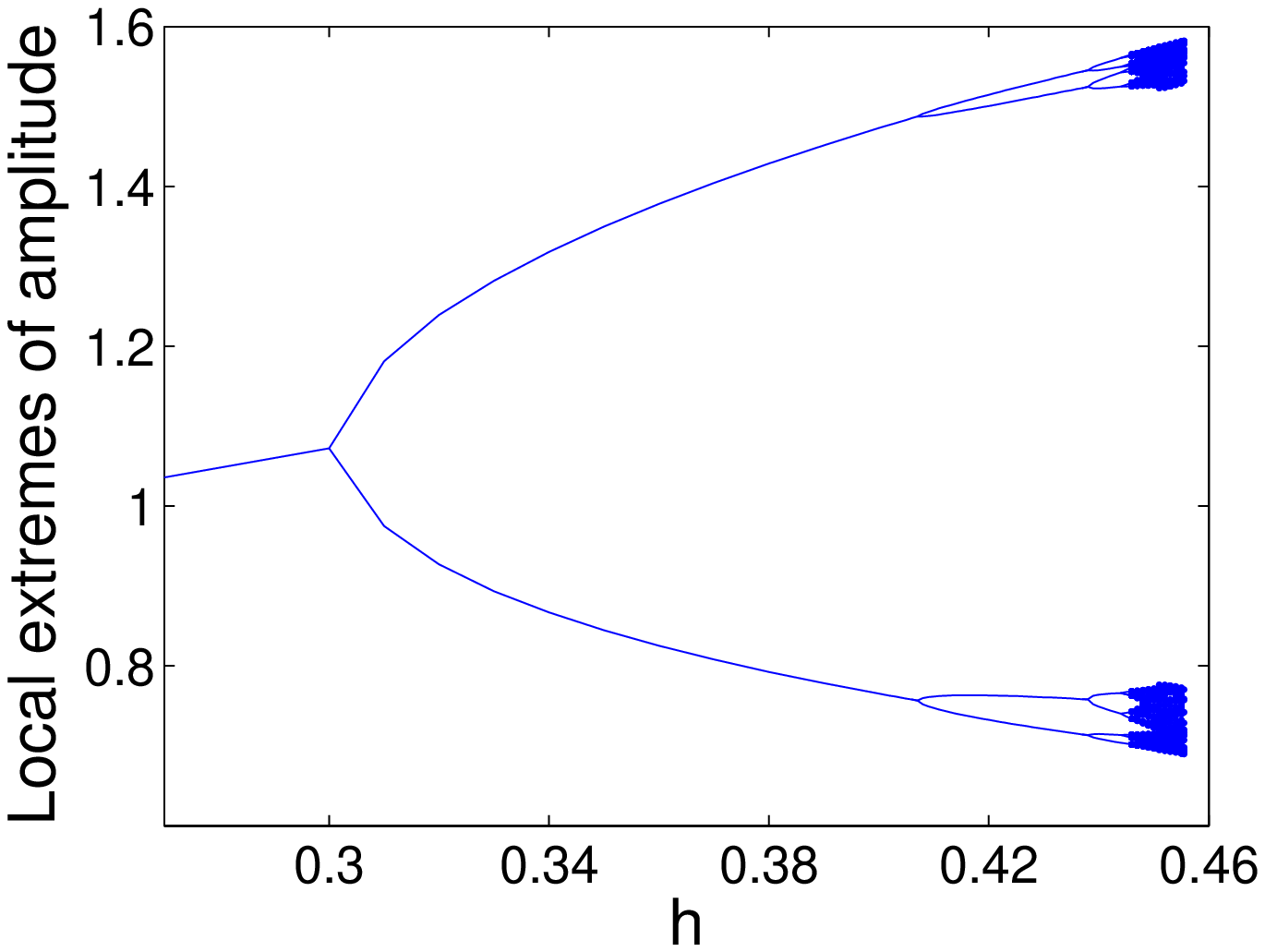}\includegraphics[scale=0.4]{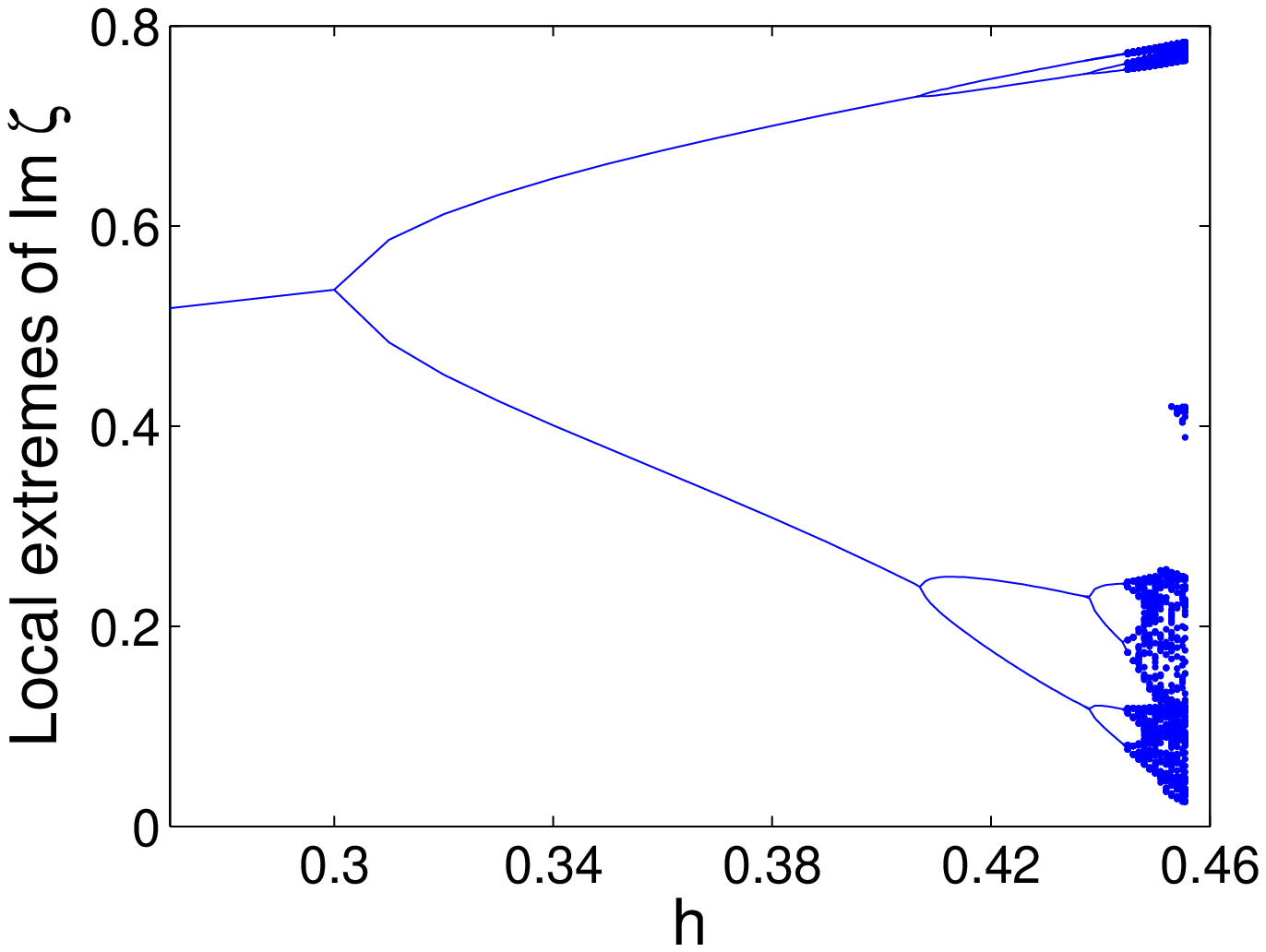}
\par\end{centering}

\begin{centering}
\emph{(a)~~~~~~~~~~~~~~~~~~~~~~~~~~~~~~~~~~~~~~~~~~~(b)}
\par\end{centering}

\caption{\emph{Period-doubling route to chaos for damping $\gamma=0.26$. The
Feigenbaum trees are shown for (a) the amplitudes and (b) ZS eigenvalues
respectively. In (c) the amplitudes of the attractor (solid line)
and the reconstructed soliton (dotted line) is shown for $h=0.4$.
The solutions (solid line) and reconstructed soliton (dotted line)
are shown for $t=2005.2$ and $t=2006.9$ in (d) and (e) respectively,
where $h=0.4.$ }\label{fig:Feigenbaum trees}}
\end{figure}

The zero attractor region is characterized by the fact that, for a
fixed damping $\gamma<0.27$, an increase in driving that leads to
the destabilization of soliton attractors is preceded by a period-doubling
route to temporally chaotic soliton attractors. To illustrate this
we consider the soliton attractors that arise when the damping is
fixed at $\gamma=0.26$.  In Figure~\ref{fig:Feigenbaum trees}~(a)
we show the local extremes of the amplitude $A_{\gamma,h}\left(t\right)$
defined in equation \eqref{eq:Amplitude}. At $h\approx0.3$ we see
that the Hopf bifurcation results in the formation of two branches,
corresponding to a 1-period soliton attractor. At $h\approx0.41$
we see the first period-doubling bifurcation, resulting in two additional
branches. On the interval $0.43<h<0.45$ we see a sequence of period-doubling
bifurcations, resulting in the formation of temporal chaos at $h\approx0.45$.
The chaos persists until $h=0.4556$ where the zero attractor region
starts. 

Barashankov and co-workers \cite{key-32} recently showed that the
chaos is \emph{caused} by a homoclinic explosion. Here we use the
soliton content to determine the \emph{effect} of these bifurcations
on radiation emissions. To this end we calculated the soliton content
associated with the attractors that arise during the period-doubling
route to temporal chaos. It is not surprising that the local extremes
of the soliton content forms a similar Feigenbaum tree. An example
is shown in Figure~\ref{fig:Feigenbaum trees}~(b) where the local
extremes of the purely imaginary ZS eigenvalue is shown for $\gamma=0.26$.
. It should be noted that for Type~II attractors we plot only the
purely imaginary ZS eigenvalue. This is justified by the fact that
side eigenvalues do not relate to the soliton part of the attractor.

A comparison between Figures~\ref{fig:Feigenbaum trees}~(a) and
(b) shows that the the shapes of the upper branches are similar. However
the shapes of the lower branches are different. In particular, the
local minima of the ZS eigenvalues appear to decrease faster than
the local minima of the attractor amplitudes. Indeed, in the former
case the ZS eigenvalues approaches zero as $h$ approaches the critical
driving strength $h=0.4556$ where the zero attractor region begins.

To analyse this we use the radiation measurement associated with the
direct scattering study. For this purpose we use the conserved quantity
of the unperturbed NLS equation to calculate the energy of radiation
emissions. In particular, the total mass (in hydrodynamics) of a potential
in the focusing NLS is defined as\begin{equation}
M\left(\psi\right)=\int_{-\infty}^{\infty}\left|\psi\right|^{2}dx.\label{eq:Total mass}\end{equation}
Inverse scattering theory allows one to separate the soliton part
and the radiation part of the mass. In particular $M\left(\psi\right)$~$=$~$M_{sol}\left(\psi\right)$~$+$~$M_{rad}\left(\psi\right)$
where the terms on the right hand side represent the total mass of
the soliton part of the solution and the total mass of the radiation
part of the soliton respectively, given by \cite{key-8}\begin{equation}
M_{sol}\left(\psi\right)=4\sum_{n=1}^{N}\eta_{n}\label{eq:Soliton mass}\end{equation}
and\begin{equation}
M_{rad}\left(\psi\right)\frac{1}{\pi}\int_{-\infty}^{\infty}\mbox{ln}\left|a\left(\xi\right)\right|^{-2}d\xi.\label{eq:Radiation mass}\end{equation}
Here the potential $\psi\left(x,0\right)$ has $N$ discrete eigenvalues
in the upper-half complex plane $\zeta_{j}=\xi_{j}+i\eta_{j}$ for
$j=1,\cdots,N$, and the function $a\left(\xi\right)$ is the reflection
coefficient that forms part of the scattering matrix (see for example
\cite{key-36}).

\begin{figure}[t]
\begin{centering}
\includegraphics[scale=0.5]{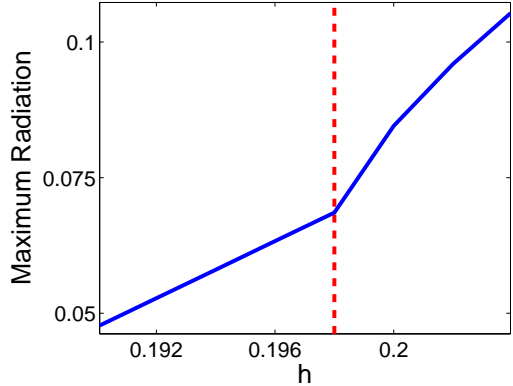}
\par\end{centering}

\caption{\emph{The maximum radiation \eqref{eq:Maximum radiation} emitted
by soliton attractors for fixed damping $\gamma=0.15$.}\label{fig:Period_doubling_bifurcation}}
\end{figure}

For attractors of the PDNLS equation, we calculated the time evolution
of the discrete eigenvalues. From equations \eqref{eq:Total mass}
and \eqref{eq:Soliton mass} the time-dependent total mass of radiation
is given by\begin{equation}
M_{rad}\left(\psi\left(x,t\right)\right)=\int_{-\infty}^{\infty}\left|\psi\left(x,t\right)\right|^{2}dx-4\sum_{n=1}^{N}\eta_{n}\left(t\right).\label{eq:Radiation mass 2}\end{equation}
Since the first term can be easily calculated numerically, we can
use equation \eqref{eq:Radiation mass 2} to measure the energy of
radiation. For Type~II soliton attractors we propose a modification
so that the side eigenvalues, associated with radiation, are excluded
from the total mass of the soliton part of the attractor. We therefore
use the following measurement of radiation mass for Type~II attractors:\begin{equation}
M_{att}\left(\gamma,h\right)=\int_{-\infty}^{\infty}\left|\psi_{\gamma,h}\left(x,t\right)\right|^{2}dx-\mbox{Im}\left\{ \zeta_{im}\left(t\right)\right\} .\label{eq:Attractor radiation mass}\end{equation}
Here $\psi_{\gamma,h}$ is the Type~II attractor that arises for
damping and driving strengths $\gamma$ and $h$ respectively, and
$\zeta_{im}\left(t\right)$ is its associated purely imaginary ZS
eigenvalue. We emphasize that this measure does not apply generally.
For example for Type~IV attractors the splitting eigenvalues would
cause a jump discontinuity. Since we are trying to establish the cause
of an imbalance that destabilizes the soliton attractor, we measure
the maximum energy of radiation. To this end we define\begin{equation}
M_{max}\left(\gamma,h\right)=\mbox{max}_{t\in\mathbb{R}}\left\{ M_{rad}\left(\psi_{\gamma,h}\left(x,t\right)\right)\right\} .\label{eq:Maximum radiation}\end{equation}

\begin{figure}[t]
\begin{centering}
\includegraphics[scale=0.6]{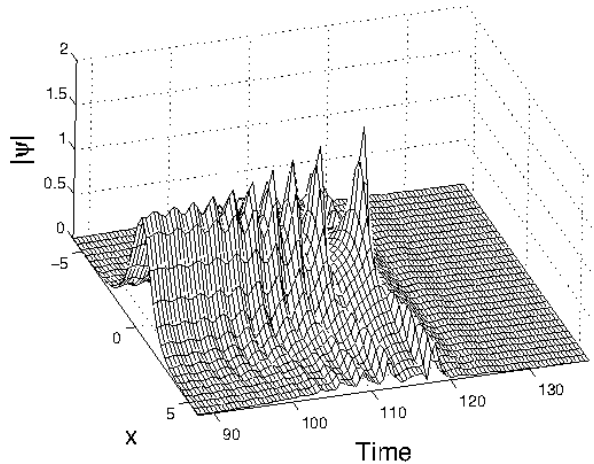}
\par\end{centering}

\emph{\caption{\emph{Soliton transient corresponding to $\gamma=0.2$ and $h=0.75$.}
\label{fig:Transient examples}}
}%
\end{figure}

From Section~3 we know that, for a fixed damping strength, an increase
in driving strength leads to an increase in radiation emissions. The
soliton content shows that the period-doubling bifurcations affect
this rate. In particular let $h_{pd}\left(\gamma\right)$ be the critical
driving strength where the first period-doubling bifurcation occurs
for a fixed damping strength $\gamma_{0}<0.27$. Numerical results
show that the following inequality holds:\begin{equation}
h_{1}<h_{pd}\left(\gamma_{0}\right)<h_{2}\Rightarrow\left.\frac{dM_{max}\left(\gamma_{0},h\right)}{dh}\right|_{h=h_{1}}<\left.\frac{dM_{max}\left(\gamma_{0},h\right)}{dh}\right|_{h=h_{2}},\label{eq:Period-doubling effect on radiation}\end{equation}
provided that $\left(\gamma_{0},h_{1}\right)$ and $\left(\gamma_{0},h_{2}\right)$
lie in the soliton attractor region. This result is illustrated in
Figure~\ref{fig:Period_doubling_bifurcation} where we show the maximum
radiation function for fixed $\gamma_{0}=0.15$. The vertical line
represents the first period-doubling bifurcation $h_{pd}\left(0.15\right)$~$\approx$~$0.198$.
It is clear that the rate of increase, i.e. the slope, is larger after
the period-doubling bifurcation.

These results show the relationship between the period-doubling route
to temporal chaos and the zero attractor region. The period-doubling
bifurcations result in a faster rate of energy losses associated with
an increase in driving. One can therefore think of the period-doubling
route to chaos as a catalyst for the formation of the zero attractor
region associated with increased driving strength.

\subsection{Soliton transient analysis for zero attractor region}

Thus far our analysis focused on oscillating solitons that arise as
attractors. For choices of damping and driving in the zero attractor
region oscillating solitons arise as transients. The study of these
transients provides insight into the role of solitons and radiation
in the transition from soliton transients to the zero attractor. In
Figure~\ref{fig:Transient examples} we illustrate a typical soliton
transient that arises for damping and driving strengths $\gamma=0.2$
and $h=0.75$ respectively. In this case we see that the soliton transient
exists until $t\approx115$. This is followed by a rapid decay to
zero. This behaviour is typical for the zero attractor region.

\begin{figure}[t]
\begin{centering}
\includegraphics[scale=0.4]{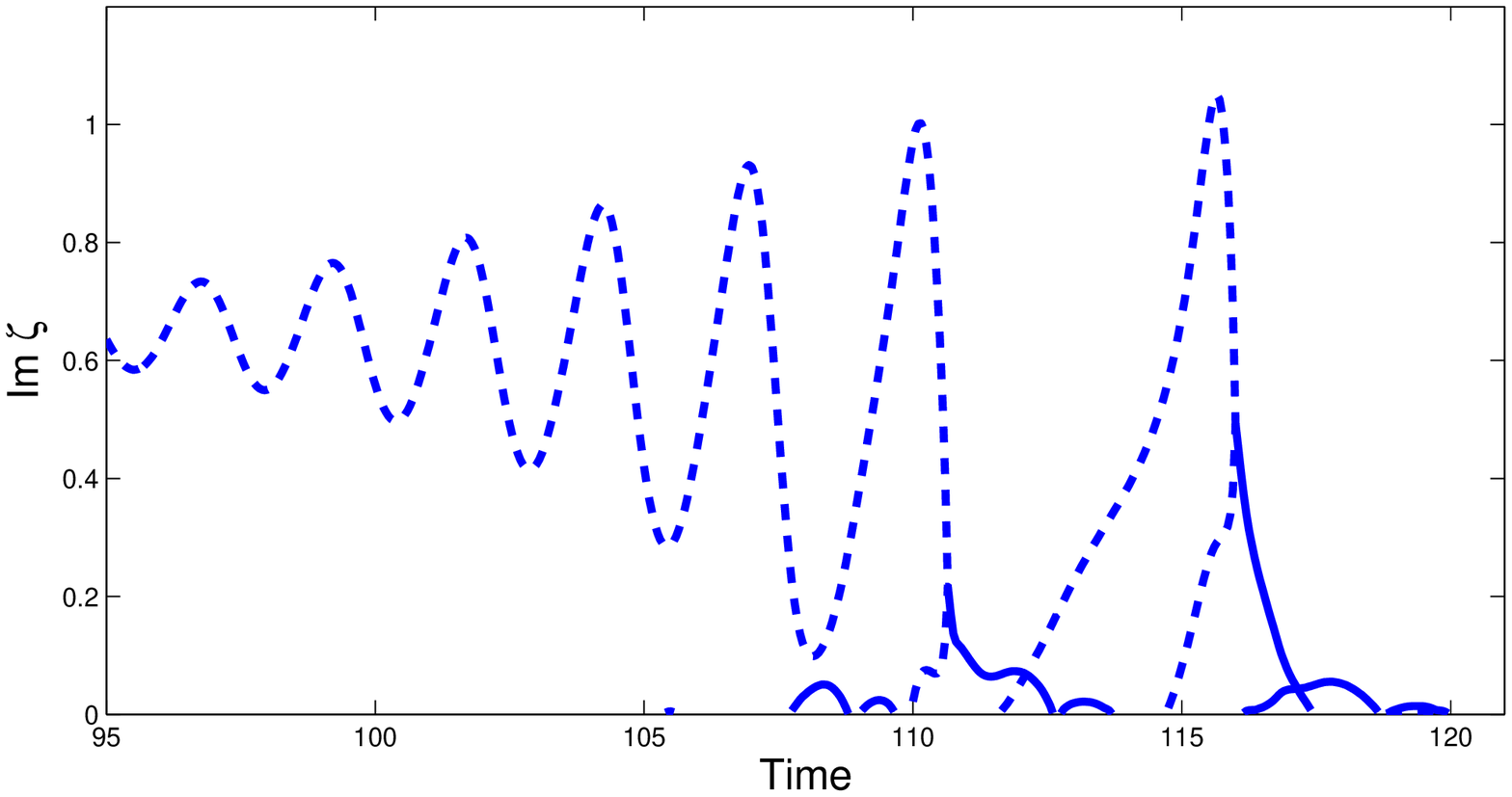}
\par\end{centering}

\begin{centering}
(a)
\par\end{centering}

\begin{centering}
\includegraphics[scale=0.4]{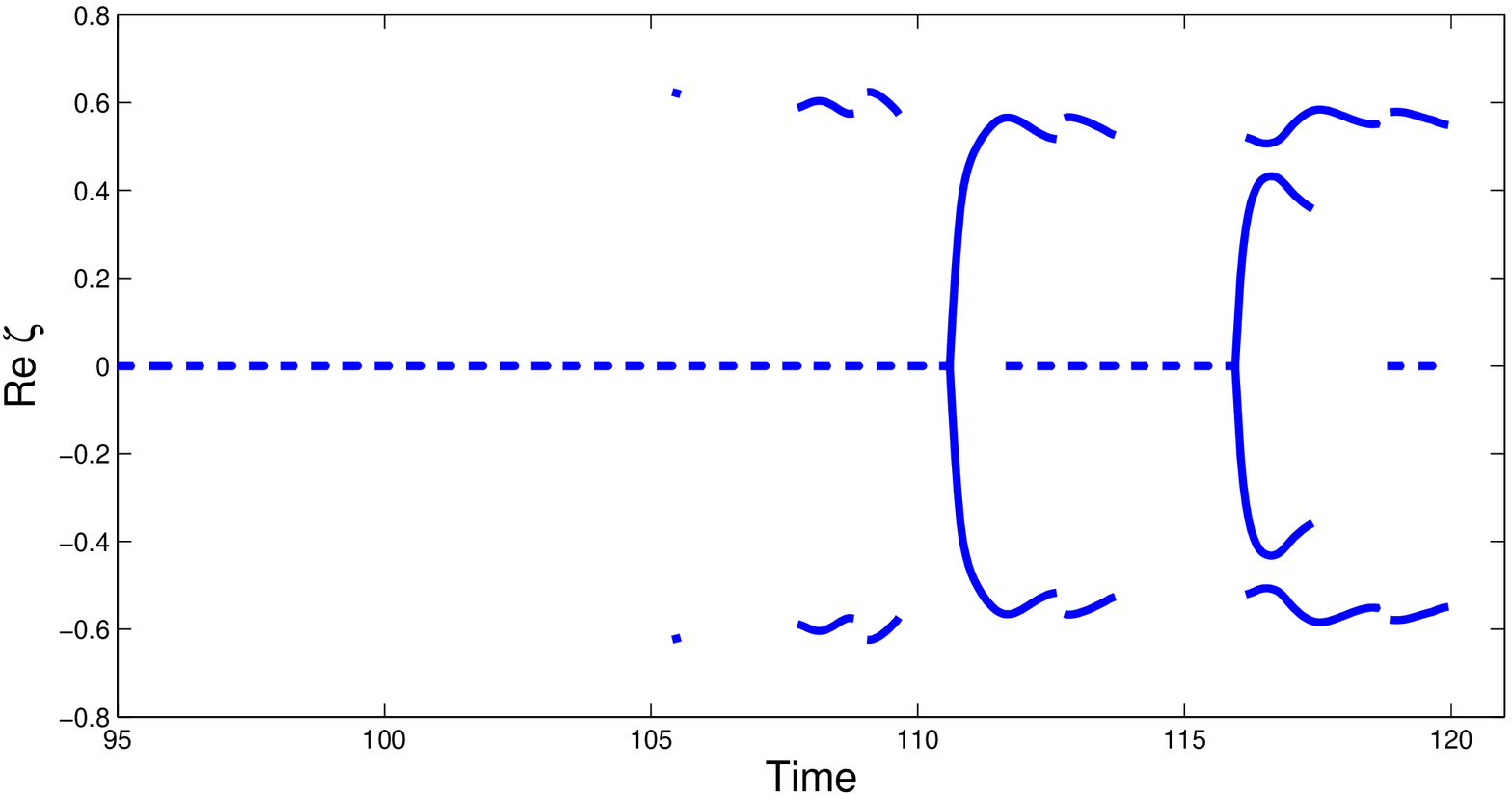}
\par\end{centering}

\begin{centering}
(b)
\par\end{centering}

\caption{\emph{The soliton content of the soliton transient arising for $\gamma=0.2$
and $h=0.75$. The imaginary and real parts of the ZS eigenvalues
are shown in (a) and (b) respectively.}\label{fig:Zero transient soliton content}}
\end{figure}

In Section~\ref{sub:Effect of period-doubling} we used the soliton
content to show that increased driving strengths lead to larger radiation
emissions that are responsible for the destabilization of the soliton
attractor. In a similar way we use the soliton content to determine
the amount of radiation required to destabilize the soliton transient.
Numerical results reveal that soliton transients behave like the four
types of soliton attractors identified in Section~\ref{sec:Direct scattering study}.
Based on this result we define different types of soliton transients.
To do this, we consider the behaviour of the soliton transient during
the final oscillation before its decay to zero. To illustrate this,
we consider the solution shown in Figure~\ref{fig:Transient examples}
corresponding to the solution associated with damping and driving
$\gamma=0.2$ and $h=0.75$ respectively. In Figure~\ref{fig:Zero transient soliton content}~(a)
and (b) we show the imaginary part and the real part of the soliton
content respectively. We see that the transient behaves like a Type~I
attractor for $t<105.$ At $t\approx105$ we see the formation of
side eigenvalues. For $105<t<110$ the solution behaves like a Type~II
attractor. However, during the final oscillation before the rapid
decay $112<t<116$ the solution behaves like a Type~IV attractor.
Based on this behaviour, the transient is classified as a Type~IV
transient.

\begin{figure}[t]
\begin{centering}
\includegraphics[scale=0.6]{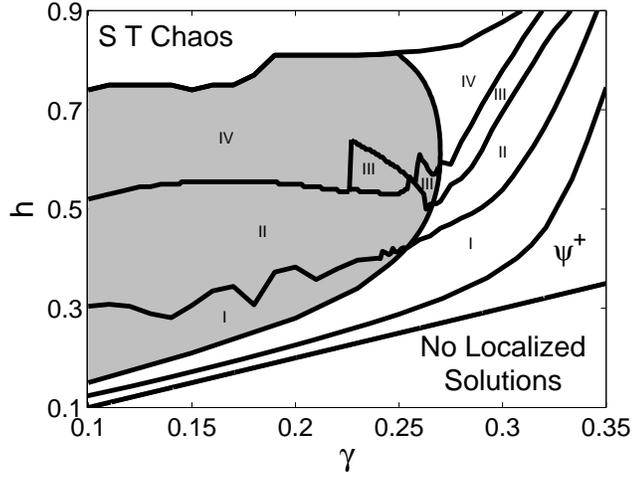}
\par\end{centering}

\centering{}\caption{\emph{Soliton transient chart. The zero attractor region is shaded.
Within this area {}``I'' shows the Type~I transients region, {}``II''
shows the Type~II transient region, {}``III'' shows the Type~III
transient region and {}``IV'' shows the Type~IV transient region.
Outside the shaded area the Roman numerals denote the different types
of attractors shown in the \ref{fig:Map} areas indicate the types
of attractors shown in Figure}~\label{fig:Transient chart}.}
\end{figure}

We calculated the types of transients throughout the zero attractor
region. The results are summarized in the transient chart Figure~\ref{fig:Transient chart}.
The shaded area corresponds to the zero attractor region. Within this
region the Roman numerals indicate the different types of transients
within this region. For example, the \emph{{}``II''} indicates the
region where Type~II transients arise. Notice that the transient
chart has the same structure as that of the soliton structure chart
Figure~\ref{fig:Map} in the intermediate damping regime, except
that the Type~III attractor region disappear for $\gamma<0.226.$
In other words, in this regime the Type~II transient region is bounded
above by the Type~IV transient region.

We can make some conclusions based on the structure of the transient
chart. The fact that the Type~I transient region is bounded above
by Type~II transients show that larger driving strengths are associated
with larger radiation emissions. The Type~II attractor region is
bounded above by either the Type~III or the Type~IV transient region,
associated with the formation of lateral waves. Once the breather-like
structure is broken, these lateral waves are emitted as radiation
waves. One can therefore conclude that even larger radiation waves
are emitted in these regions. One can therefore conclude that an increase
in driving strength has two opposite effects on oscillating solitons.
On the one hand, it increases the magnitude of temporal oscillations,
resulting in larger radiation emissions. This has a destabilization
effect that results in the formation of the zero attractor region.
On the other hand it sustains the soliton in the sense that larger
radiation emissions are required to destroy the soliton transient.
The latter effect is responsible for the restabilization of the soliton
attractor $\left(0.25\leq\gamma\leq0.27\right)$ and the formation
of spatio-temporal chaos $\left(\gamma<0.25\right)$.

\section{The spatio-temporal chaotic region\label{sec:5}}

\begin{center}
\begin{figure}[t]
\begin{centering}
\includegraphics[scale=0.5]{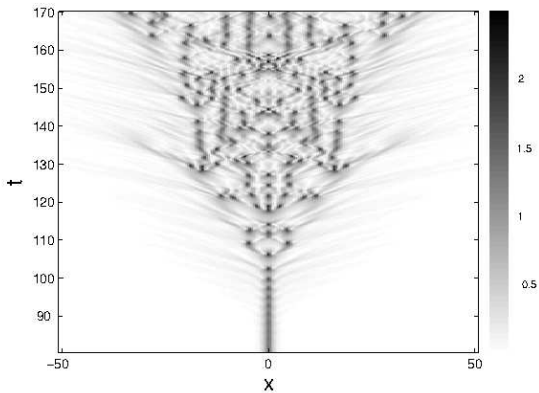}
\par\end{centering}

\caption{\emph{Onset of chaos for $\gamma=0.2$ and $h=0.82$.}\label{fig:Spatio-temporal chaos map and example}}
\end{figure}

\par\end{center}

Spatio-temporal chaos forms as a result of exceedingly large driving
strengths. When $\psi^{+}$ is integrated in this region a soliton
transient emerges. At a critical point in time additional solitons
start to form. This moment marks the onset of spatio-temporal chaos.
As time evolves beyond this point more solitons are formed, resulting
in the solution to spread across the spatial domain. A typical example
of the formation of chaos is shown in Figure~\ref{fig:Spatio-temporal chaos map and example}.
Here we integrated $\psi^{+}$ for $\gamma=0.2$ and $h=0.82.$ For
$t<105$ we see a soliton transient. At $t\approx108$ we see the
formation of two solitons, marking the onset of spatio-temporal chaos.
Beyond $t\approx108$ we see that the formation of more solitons results
in the the spreading of the solution across the spatial domain.

\begin{figure}[t]
\noindent \begin{centering}
\includegraphics[scale=0.4]{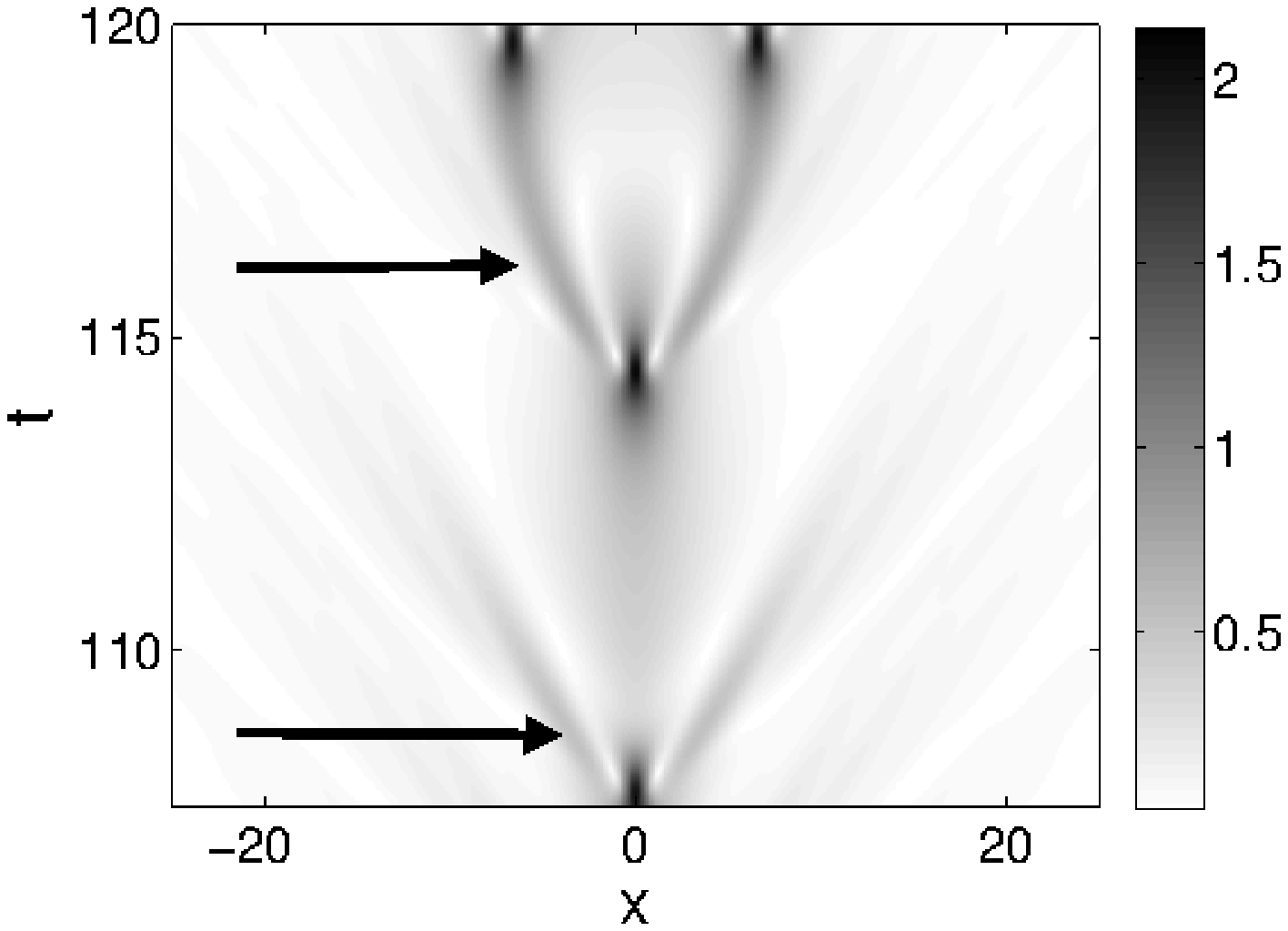}
\par\end{centering}

\begin{centering}
(a)
\par\end{centering}

\noindent 
\begin{centering}
\includegraphics[scale=0.4]{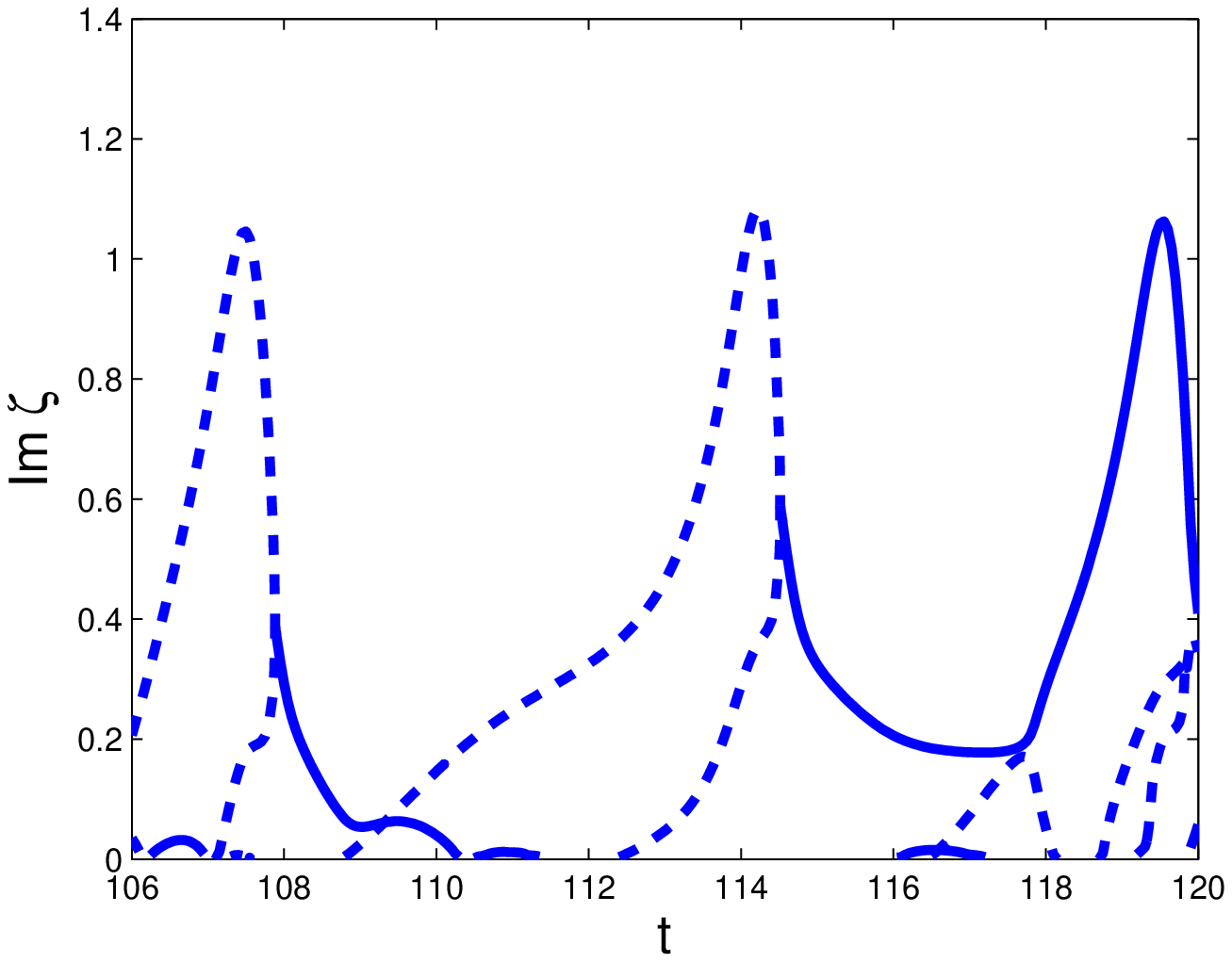}\includegraphics[scale=0.4]{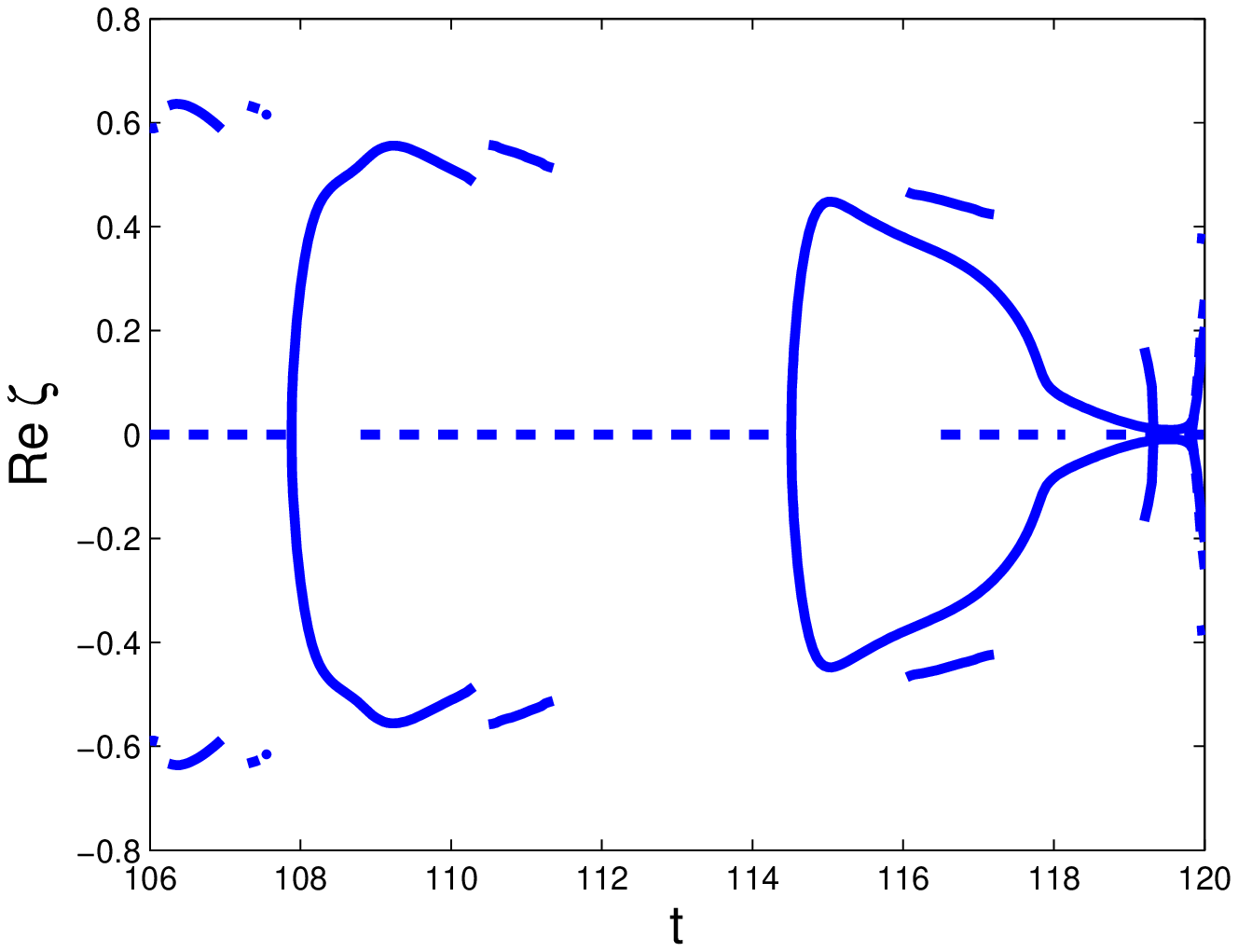}
\par\end{centering}

\begin{centering}
(b)~~~~~~~~~~~~~~~~~~~~~~~~~~~~~~~~~~~~~~~~~~~~~~(c)
\par\end{centering}

\caption{\emph{Onset of chaos in small damping regime arising for $\gamma=0.2$
and $h=0.802$. In (a) the solution is shown. The imaginary and real
parts of the associated ZS eigenvalues} \emph{are shown in (b) and
(c) respectively.}\label{fig:Formation of chaos small damping}}
\end{figure}

To investigate the transition from soliton transients to spatio-temporal
chaos we integrate $\psi^{+}$ numerically for damping and driving
strengths near the lower boundary of the spatio-temporal chaotic region.
We also calculate the soliton content during the transition from soliton
transient to spatio-temporal chaos in order to monitor the role of
the solitons and radiation in this transition.

Numerical results show that, depending on the damping strength, either
lateral waves or radiation tails are responsible for the formation
of additional solitons. In particular, for smaller damping strengths
$\gamma\leq0.34$ lateral waves are responsible for the formation
of additional solitons, leading to chaos. In this case the lateral
waves become solitons. Note that this is in contrast to the Type~IV
attractors where lateral waves become radiation waves. For larger
damping strengths $\gamma>0.34$ the formation of additional solitons
is associated with radiation tails, consisting of multiple radiation
waves, that couple to form additional solitons. In the latter case
the role of the soliton is to {}``feed'' the radiation tail, resulting
in the formation of solitons in the tail of the soliton. We refer
to the role of the soliton in the smaller and larger damping regime
as direct and indirect respectively.

The direct role of the soliton in the formation of spatio-temporal
chaos for the smaller damping regime is illustrated in Figure~\ref{fig:Formation of chaos small damping}
where the results are shown for damping strength $\gamma=0.2$, and
driving strength $h=0.802$. In Figure~\ref{fig:Formation of chaos small damping}~(a)
the transition from soliton transient to spatio-temporal chaos is
shown. The lower arrow shows the lateral wave that forms during the
penultimate temporal oscillation. One can clearly see that this wave
dissipates harmlessly, and plays no role in the formation of additional
solitons. The upper arrow shows the lateral wave that form during
the final temporal oscillation before the formation of multiple solitons.
We see that this wave is transformed from lateral wave to soliton
at $t\approx120$. In Figure~\ref{fig:Formation of chaos small damping}~(b)
and (c) we show the imaginary and real parts of the associated ZS
eigenvalues respectively. Notice that splitting eigenvalues are observed
during the formation of multiple solitons on the interval $115\leq t\leq120$.
Splitting eigenvalues are observed whenever the lateral waves are
responsible for the formation of additional solitons. This shows that
the formation of lateral waves in the breather-like structures of
Type~III and Type~IV attractors are responsible for the formation
of chaos in the smaller damping regime.

\begin{figure}[t]
\begin{centering}
\includegraphics[scale=0.4]{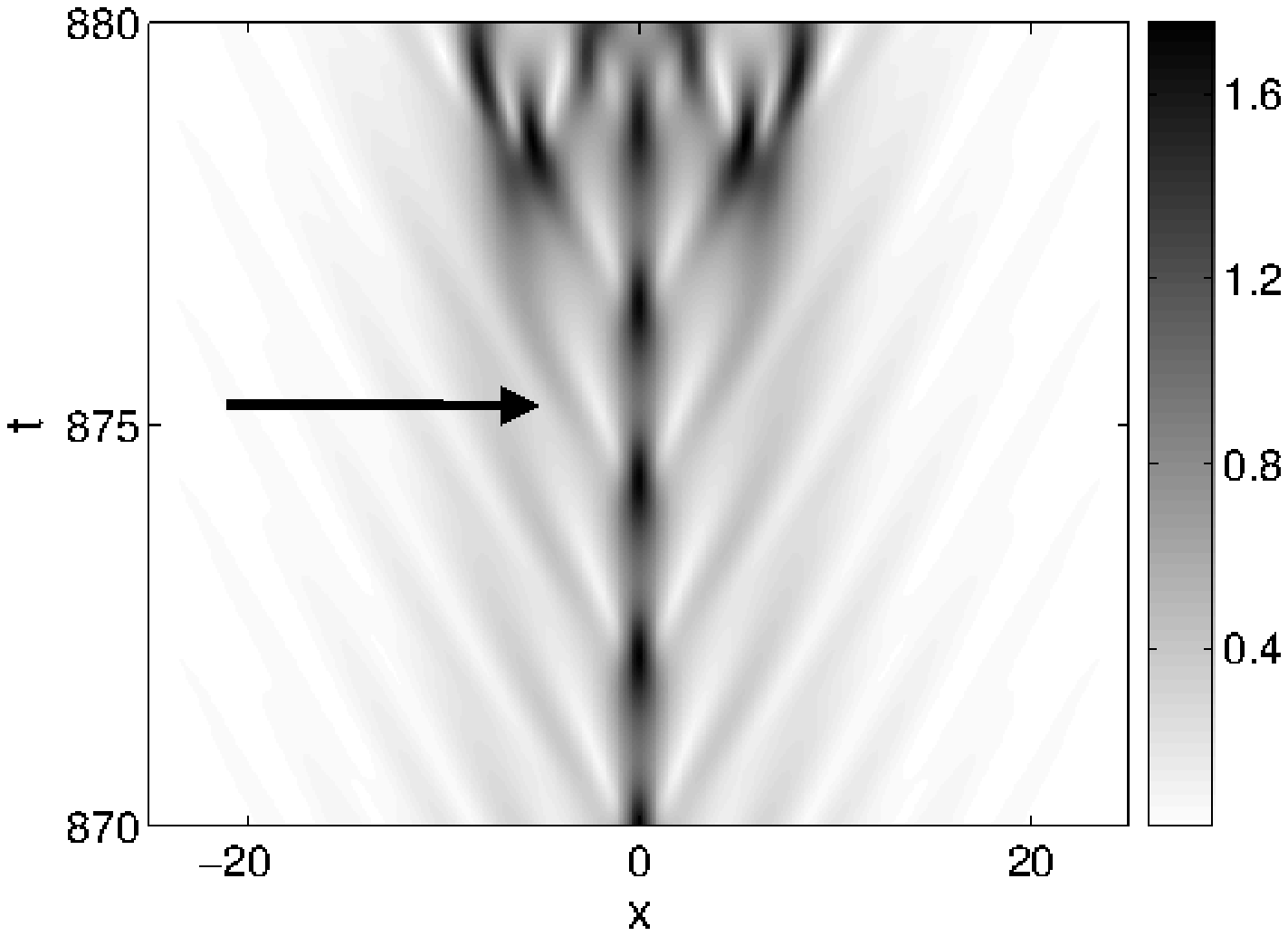}
\par\end{centering}

\begin{centering}
(a)
\par\end{centering}

\begin{centering}
\includegraphics[scale=0.4]{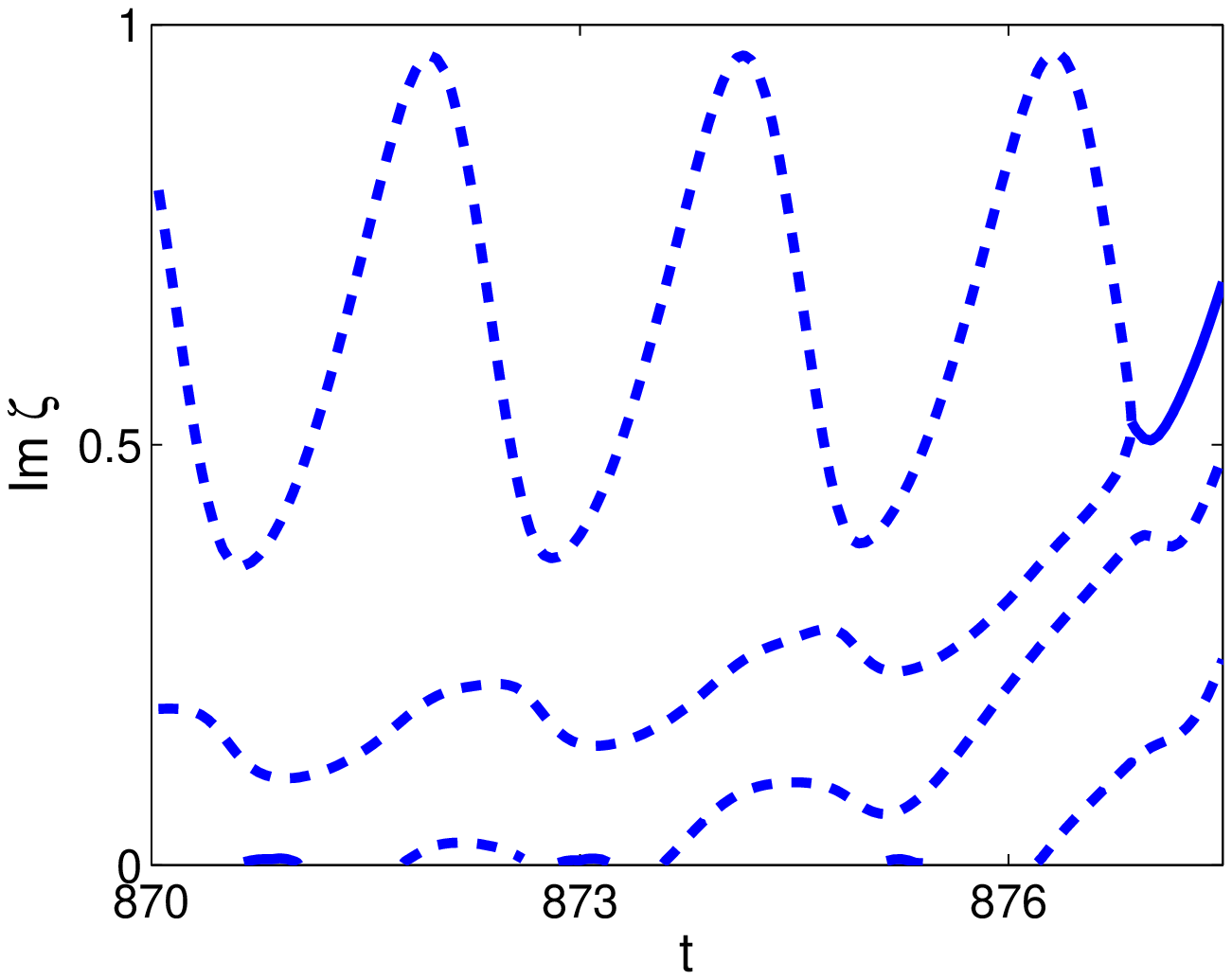}\includegraphics[scale=0.4]{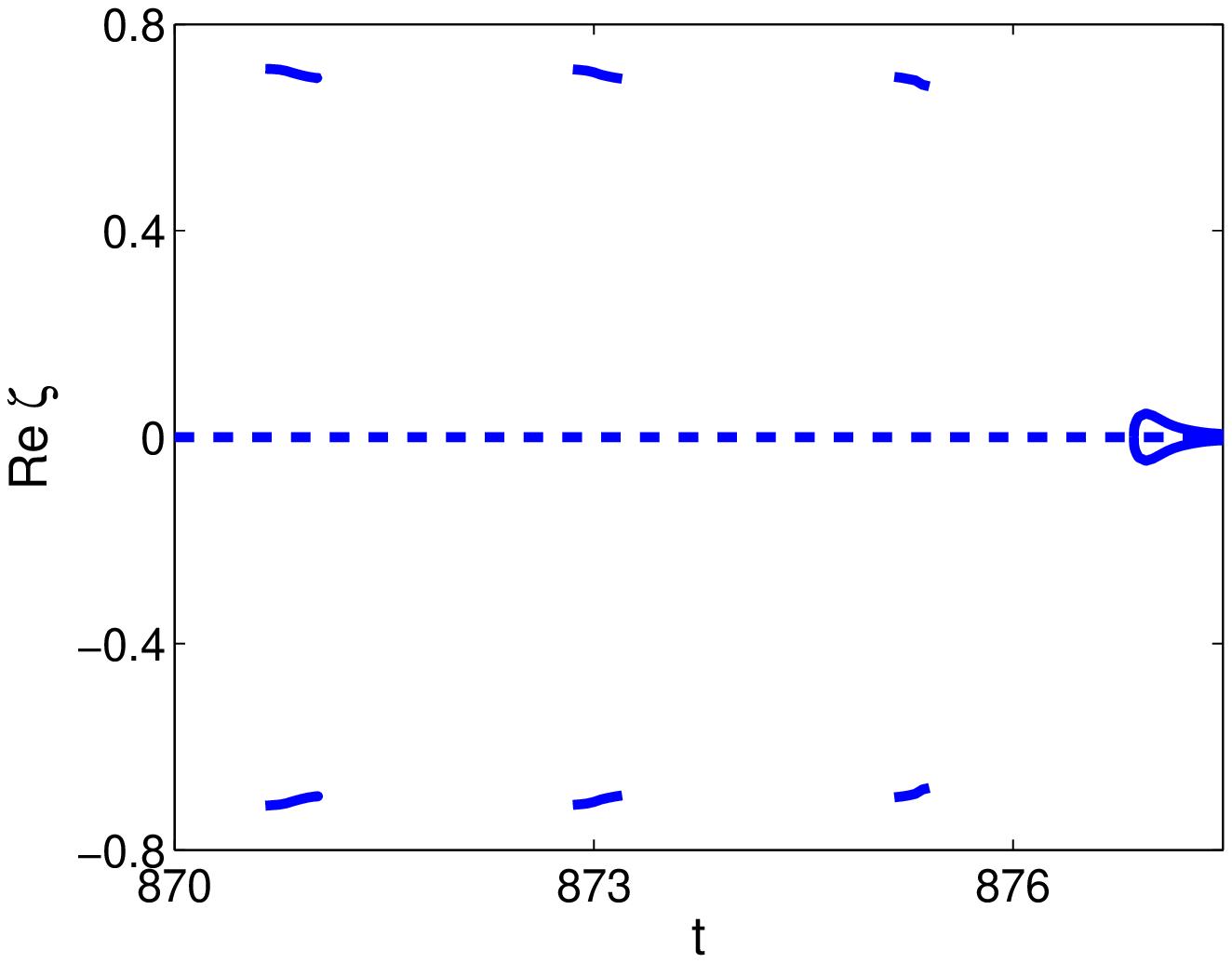}
\par\end{centering}

\begin{centering}
(b)~~~~~~~~~~~~~~~~~~~~~~~~~~~~~~~~~~~~~~~~~~~~~~(c)
\par\end{centering}

\caption{\emph{Onset of chaos in larger damping regime for $\gamma=0.35$ and
$h=1.019$. In (a) the solution is shown. The imaginary and real parts
of the associated ZS eigenvalues} \emph{are shown in (b) and (c) respectively.}\label{fig:Formation of chaos medium damping}}
\end{figure}

\begin{figure}[t]
\noindent \begin{centering}
\includegraphics[scale=0.4]{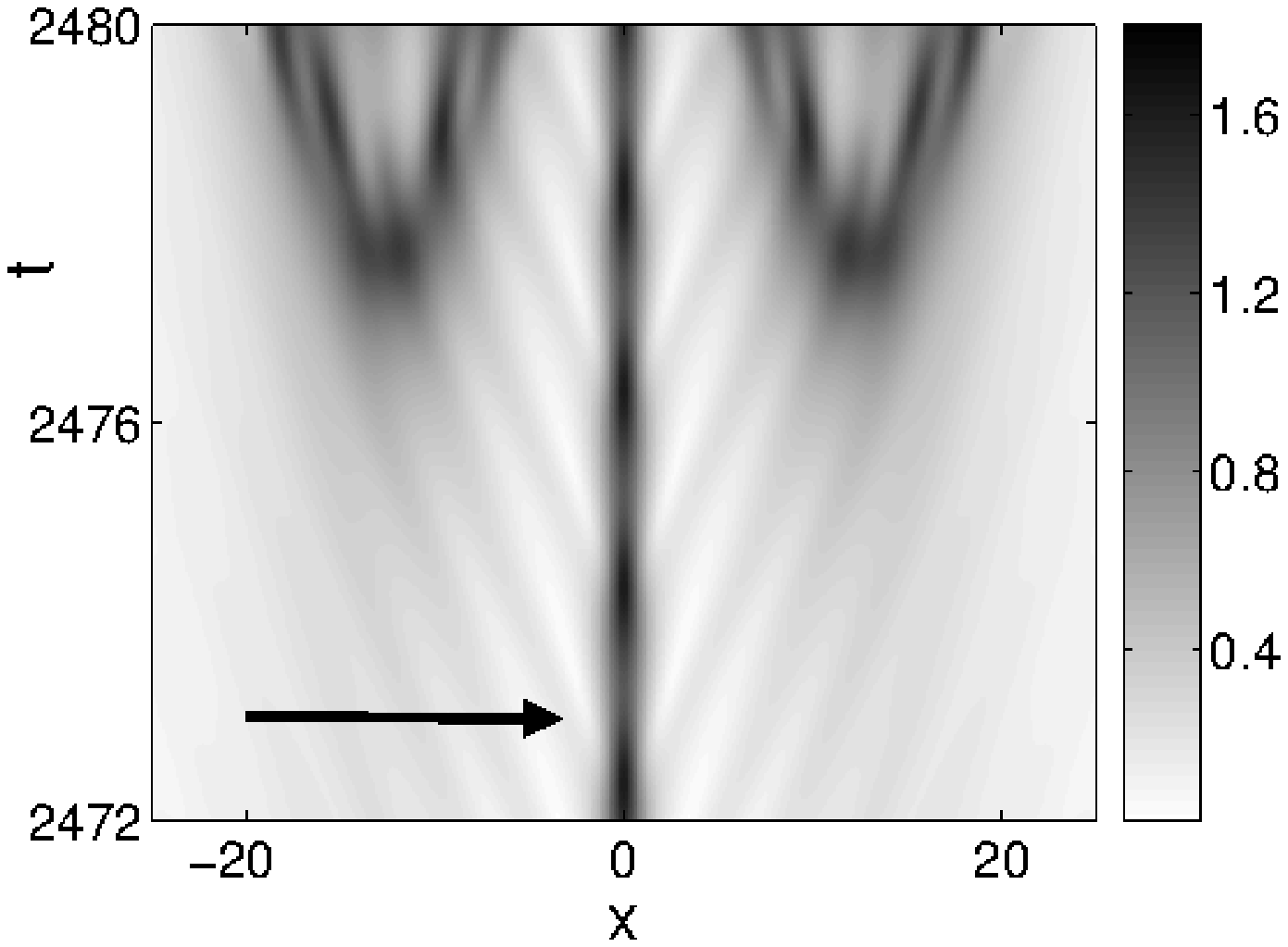}
\par\end{centering}

\begin{centering}
(a)
\par\end{centering}

\begin{centering}
\includegraphics[scale=0.4]{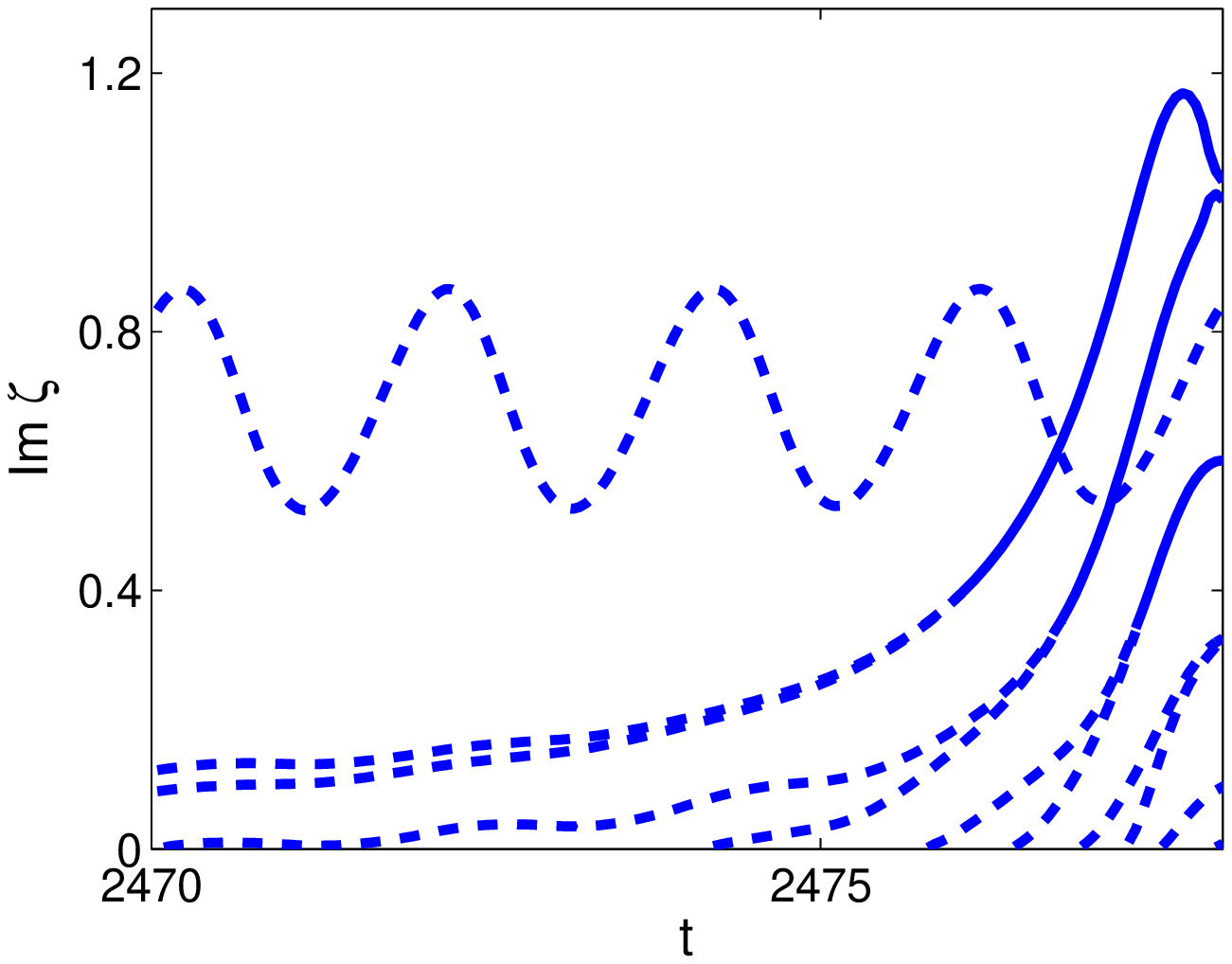}\includegraphics[scale=0.4]{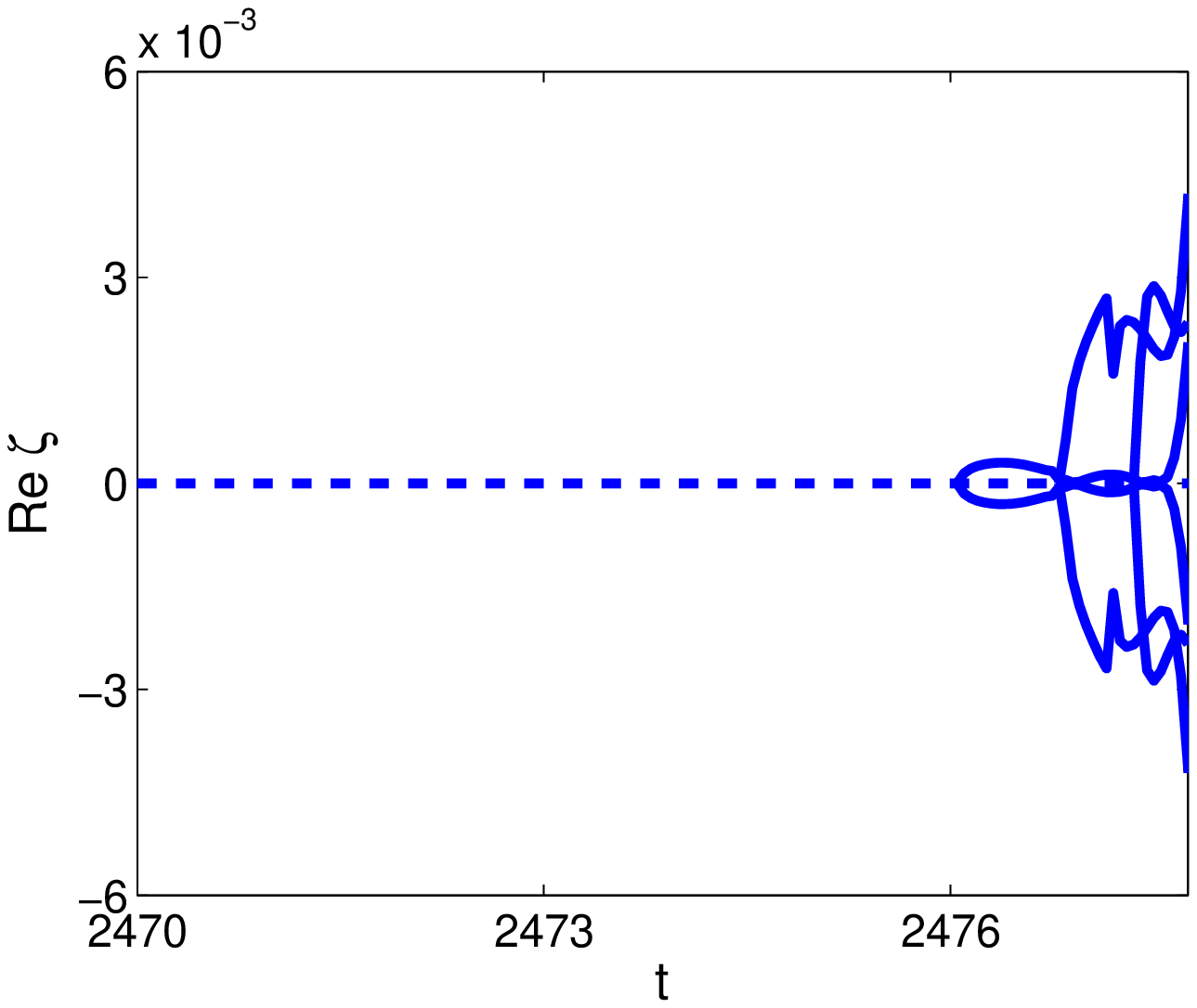}
\par\end{centering}

\begin{centering}
(b)~~~~~~~~~~~~~~~~~~~~~~~~~~~~~~~~~~~~~~~~~~~~~~~(c)
\par\end{centering}

\caption{\emph{Onset of chaos in larger damping regime for $\gamma=0.37$ and
$h=1.063$. In (a) the solution is shown. The imaginary and real parts
of the associated ZS eigenvalues} \emph{are shown in (b) and (c) respectively.}\label{fig:Formation of chaos large damping}}
\end{figure}

The indirect role of the soliton in the formation of spatio-temporal
chaos for the larger damping regime is illustrated in Figures~\ref{fig:Formation of chaos medium damping}
and \ref{fig:Formation of chaos large damping}, corresponding to
damping strengths $\gamma=0.35$ and $\gamma=0.37$ respectively.
In Figure~\ref{fig:Formation of chaos medium damping}~(a) we show
the solution that arises when $\psi^{+}$ is integrated numerically
for $\gamma=0.35$ and $h=1.019$. The arrow shows the lateral wave
formed during the penultimate temporal oscillation of the soliton
transient. At $t\approx879$ we see that this wave combines with another
radiation wave (formed during the final temporal oscillation of the
transient) to form additional solitons. In Figure~\ref{fig:Formation of chaos medium damping}~(b)
and (c) we show the imaginary and real parts of the associated ZS
eigenvalues respectively. In the former we see that no splitting occurs
until the formation of additional solitons at $t\approx877.$ Figure~\ref{fig:Formation of chaos medium damping}~(c)
shows that the soliton content of the soliton transient contains side
eigenvalues. In Figure~\ref{fig:Formation of chaos large damping}~(a)
we show the solution that arises when $\psi^{+}$ is integrated numerically
for damping and driving strengths $\gamma=0.37$ and $h=1.063$ respectively.
The arrow shows the radiation wave that is emitted at $t\approx2472$,
three oscillations before the formation of multiple solitons. Here
we see that this radiation wave does not dissipate, but remains in
the tail of the soliton. Finally at $t\approx2477$ this wave combines
with other radiation waves to form multiple solitons. In (b) and (c)
we show the real and imaginary parts of the associated ZS eigenvalues
respectively. In this case we see no splitting eigenvalues. Instead
we see the formation of multiple purely imaginary ZS eigenvalues.
This is typical of the transition from the soliton transient to chaos
when radiation tails are responsible for the formation of multiple
solitons.

We conclude that the soliton part of the soliton transient plays a
direct role in the formation of spatio-temporal chaos for the damping
regimes $\gamma\leq0.34$. The splitting eigenvalues show that the
soliton breaks up to form lateral waves. The lateral waves are sufficiently
large so that the effect of driving dominates the dissipative effects.
The result is that the lateral waves become solitons that lead to
the formation of spatio-temporal chaos. In contrast, in the large
damping regime $0.34<\gamma\leq0.372$ the soliton part of the soliton
transient plays a more indirect role in the sense that its role is
restricted to the emission of radiation waves. In this setting the
strong driving strength weakens the effect of dissipation, so that
the lifetime of radiation waves increase. These radiation waves interact,
resulting in the formation of multiple solitons.

It should be noted that these results tie in well with the soliton
structure chart Figure~\ref{fig:Map}, where the spatio-temporal
chaotic attractor region is bounded below by Type~IV attractors for
$0.24<\gamma<0.328$. We also know from the transient chart Figure~\ref{fig:Transient chart}
that the spatio-temporal chaotic attractor region is bounded below
by Type~IV transients for $\gamma<0.24$. These results can be interpreted
as follows: Chaos in the regime $\gamma<0.328$ where the spatio-temporally
chaotic region is bounded below by a Type~IV attractor/transient
region is caused by lateral waves. Chaos in the regime $\gamma>0.351$
where the spatio-temporally chaotic region is bounded below by Type~I
or Type~II attractors indicate that, at increased driving strengths,
radiation tails are responsible for the formation of spatio-temporal
chaos. Chaos in the regime $0.328\leq\gamma\leq0.351$ where the spatio-temporally
chaotic region is bounded below by Type~III attractors are formed
either by lateral waves ($\gamma\leq0.34$) or radiation wave trains
($\gamma>0.34$).

\section{Conclusions}

In this paper we used the direct scattering study to investigate the
dynamics of the parametrically driven nonlinear Schr\"odinger equation.
We used the scattering data in two ways, namely soliton identification
and radiation emissions measurement. The former were used to identify
four different types of time-dependent soliton attractors, each with
a unique soliton structure. The results show that the soliton attractors
associated with moderate driving strengths (near the Hopf bifurcation
boundary) consist of only a single nonlinear mode. The increase of
radiation emissions associated with increased driving strengths lead
to the temporary excitation of additional nonlinear modes in the form
of side eigenvalues. In the intermediate damping regime large driving
strengths lead to the formation of breather-like structures, characterised
by the formation of a lateral wave on each side of the soliton. Larger
driving strengths lead to the destruction of the soliton structure
on a periodic basis, resulting in the loss of nonlinear modes. This
is accompanied by the periodic formation of additional nonlinear modes.

Radiation emission measurements were used to investigate the transition
from period-doubling bifurcations to temporal chaos to destabilization
of soliton attractors associated with increased driving strengths.
The calculation of radiation emissions show that the rate of radiation
emission, associated with increased driving strengths, increases faster
when period-doubling bifurcations occur. This suggests that period-doubling
may act as a catalyst for the destabilization of the soliton attractor
that results in the zero attractor region. 

We also used the direct scattering study to analyse soliton transients.
For the zero attractor region we showed that soliton transients have
the same structure as soliton attractors. The results show that larger
radiation emissions are required to destroy soliton transients for
larger driving strengths. This is evident from the restabilization
that occurs for the damping regime $0.25<\gamma<0.27.$ A study of
the soliton transients that arise for the spatio-temporally chaotic
region showed that the formation of breather-like structures, associated
with Type~III and Type~IV attractors, play an essential role in
the formation of chaos for the damping regime $\gamma\leq0.34$. For
the larger damping regime $\gamma>0.34$ chaos forms as a result of
coupled radiation waves in the radiation tail.

\section{Acknowledgments}

CO would like to thank the National Institute of Theoretical Physics
(NITheP) and the University of Cape Town's postgraduate funding office and PPI fund
for financial assistance. He would also like to thank the University
of Cape Town's ICTS High Performance Computing Team for support with the computer
simulations.

\section*{Appendix: The soliton content for the unperturbed NLS equation}

Here we consider the relationship between the soliton content and
the associated soliton. This is done by reconstructing soliton solutions
for three different pairs of ZS eigenvalues namely \emph{1)} a single
ZS eigenvalue, \emph{2)} two purely imaginary ZS eigenvalues and \emph{3)}
two complex ZS eigenvalues symmetric about the imaginary axis. For
each case we use the reconstructed solitons to identify the behaviour
associated with the soliton content of the four types of attractors
reported in Section~\ref{sec:Direct scattering study}.

\subsubsection*{Single ZS eigenvalue}

Potentials corresponding to a single ZS eigenvalue $\zeta=\xi+i\eta$
are associated with solitons of the form \begin{equation}
q\left(x,t\right)=2\eta\mbox{sech}\left[2\eta\left(x+4\xi t\right)-2\delta_{0}\right]\mbox{exp}\left[-2i\xi x+4i\left(\eta^{2}-\xi^{2}\right)t-i\left(\psi_{0}+\frac{\pi}{2}\right)\right].\label{eq:Single soliton}\end{equation}
These solitons are bell-shaped waves propagating at a constant velocity.
From the solution \eqref{eq:Single soliton} it follows that the amplitude
of the soliton is given by $A=2\eta$, while the velocity is given
by $V=-4\xi$, and the width of the soliton is inversely proportional
to the amplitude $A$. Therefore, in cases where the discrete spectrum
consists of a single ZS eigenvalue, the soliton content reveals that
the solution contains only a single soliton whose amplitude and width
depend on the imaginary part of the ZS eigenvalue, while its velocity
is proportional to the real part of the ZS eigenvalue.

It is important to note that the soliton content does not contain
the initial position and initial phase $\delta_{0}$ and $\psi_{0}$
respectively. To reconstruct these quantities, one has to obtain the
normalisation coefficient additionally. This falls outside the scope
of our application of the direct scattering study.

\subsubsection*{Two purely imaginary ZS eigenvalues}

\begin{figure}[t]
\begin{centering}
\includegraphics[scale=0.4]{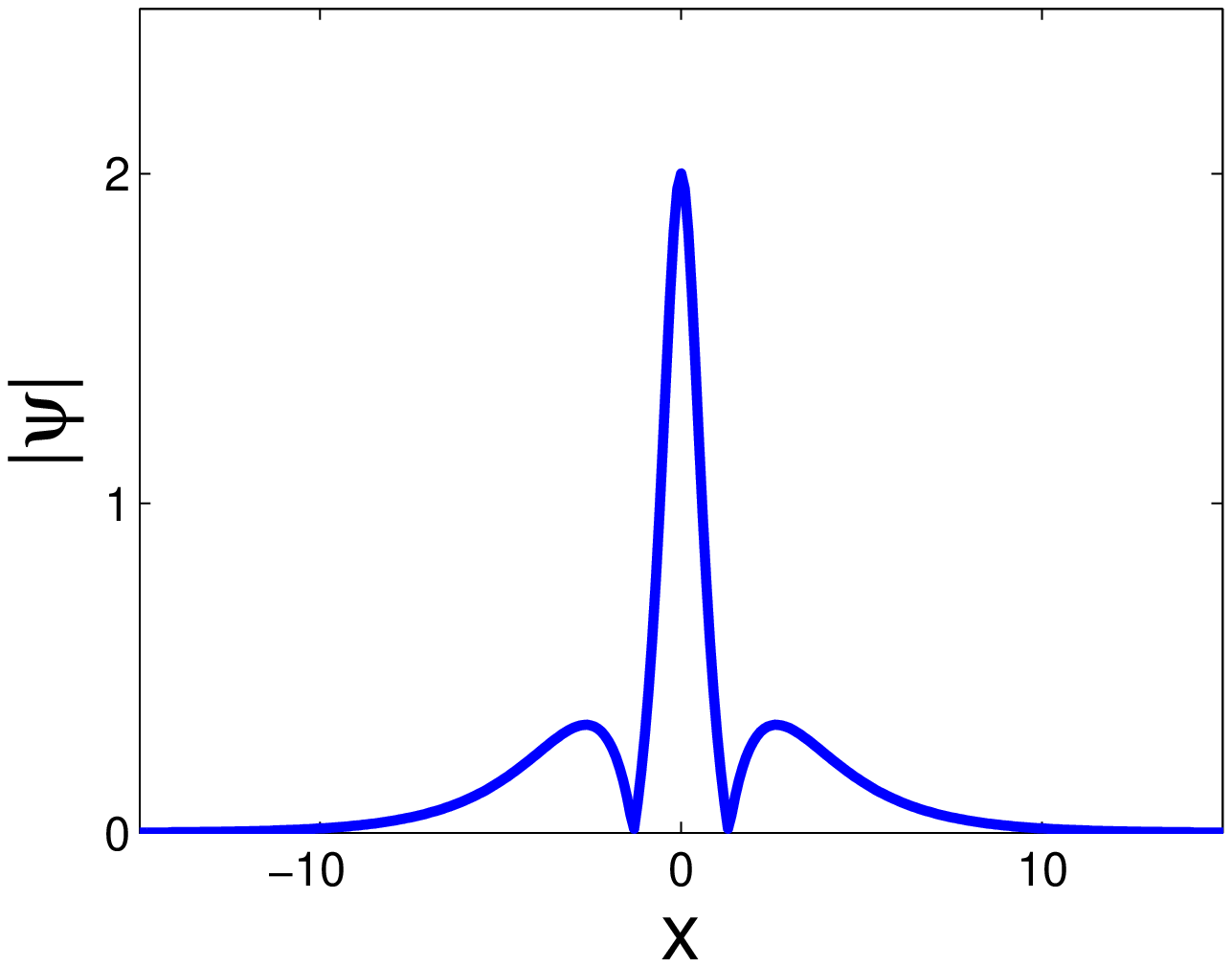}~~~~~\includegraphics[scale=0.4]{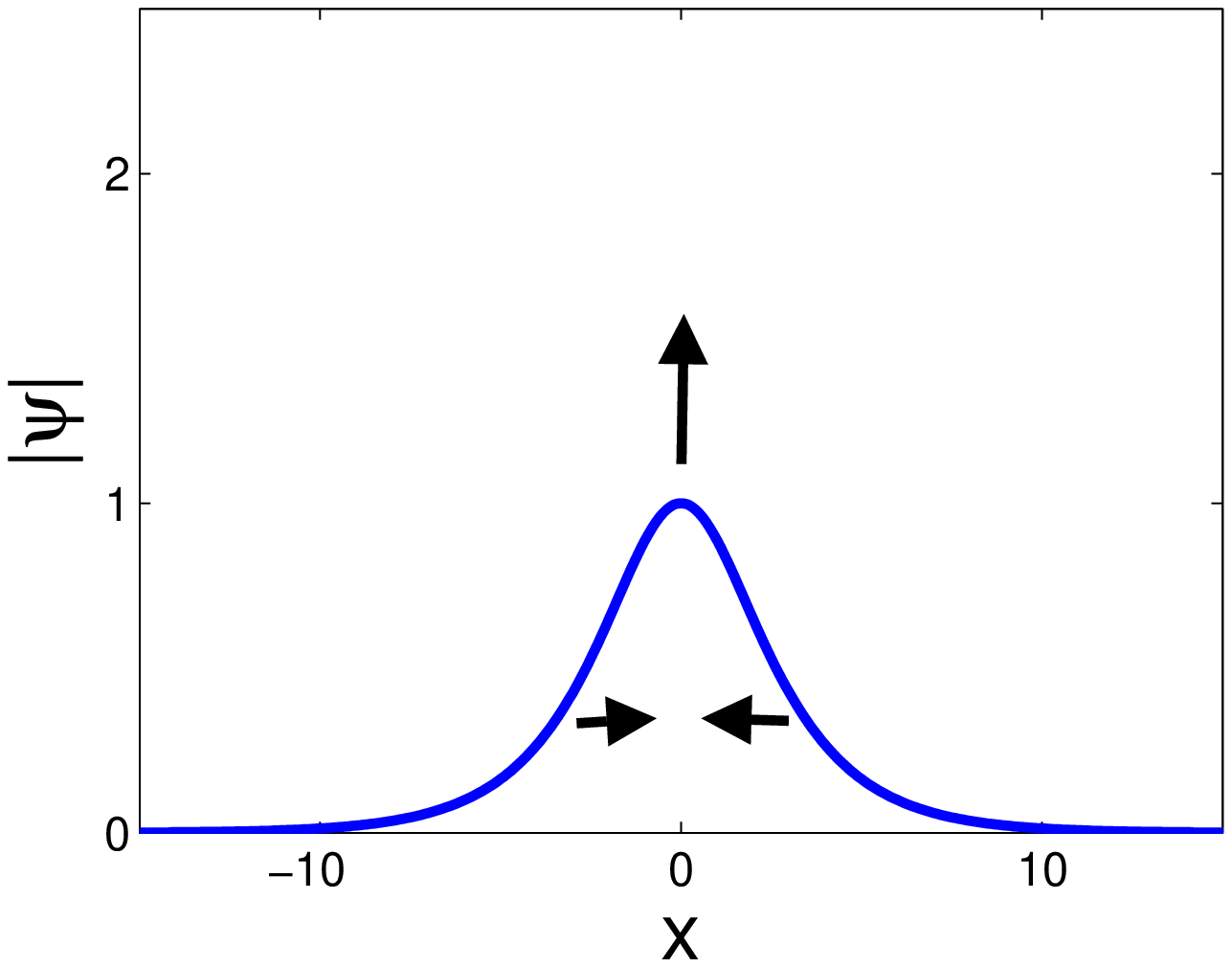}
\par\end{centering}

\begin{centering}
(a)\qquad{}\qquad{}\qquad{}\qquad{}\qquad{}\qquad{}\qquad{}~~~~~~~~~~(b)
\par\end{centering}

\begin{centering}
\includegraphics[scale=0.4]{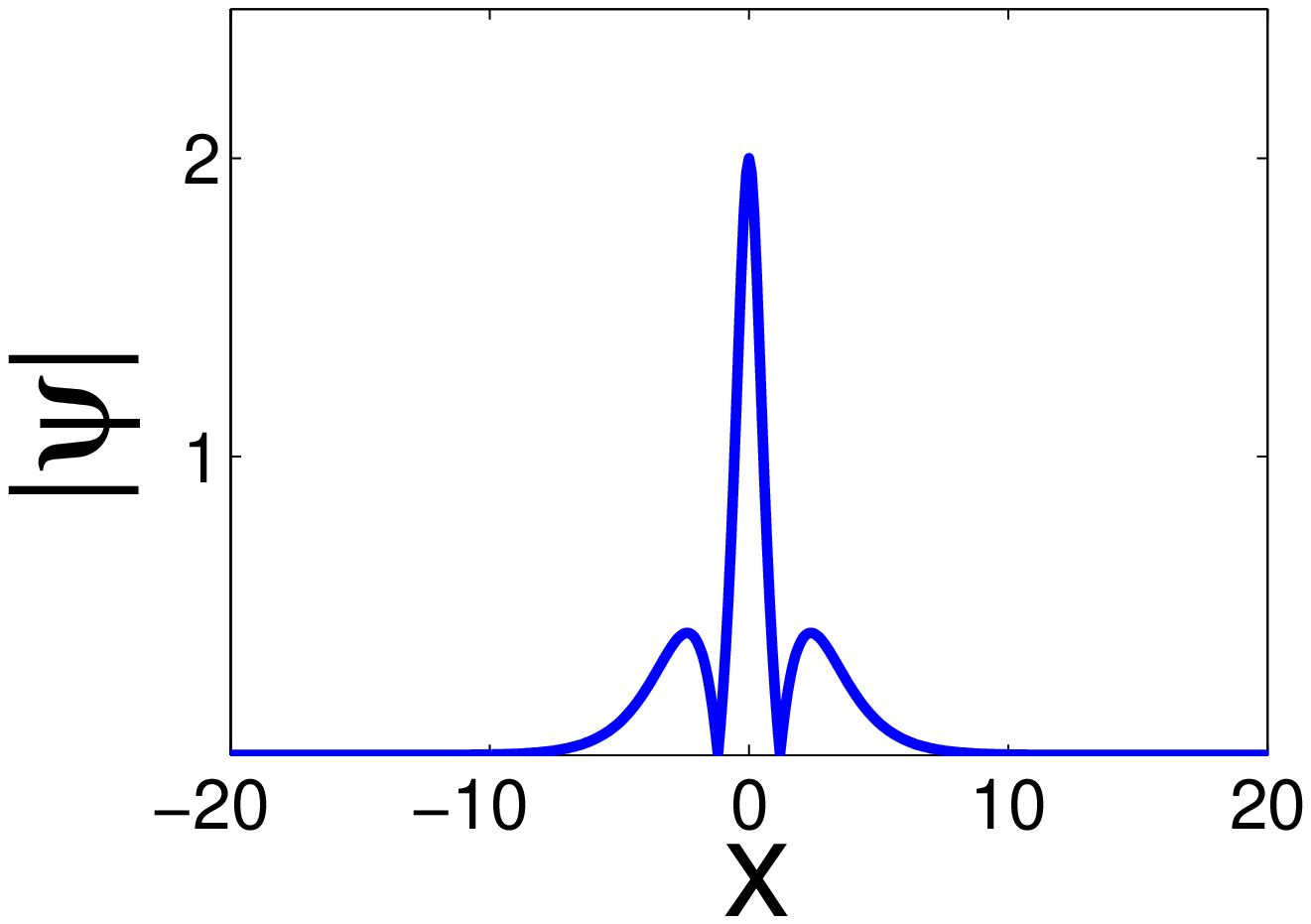}~~~~~\includegraphics[scale=0.4]{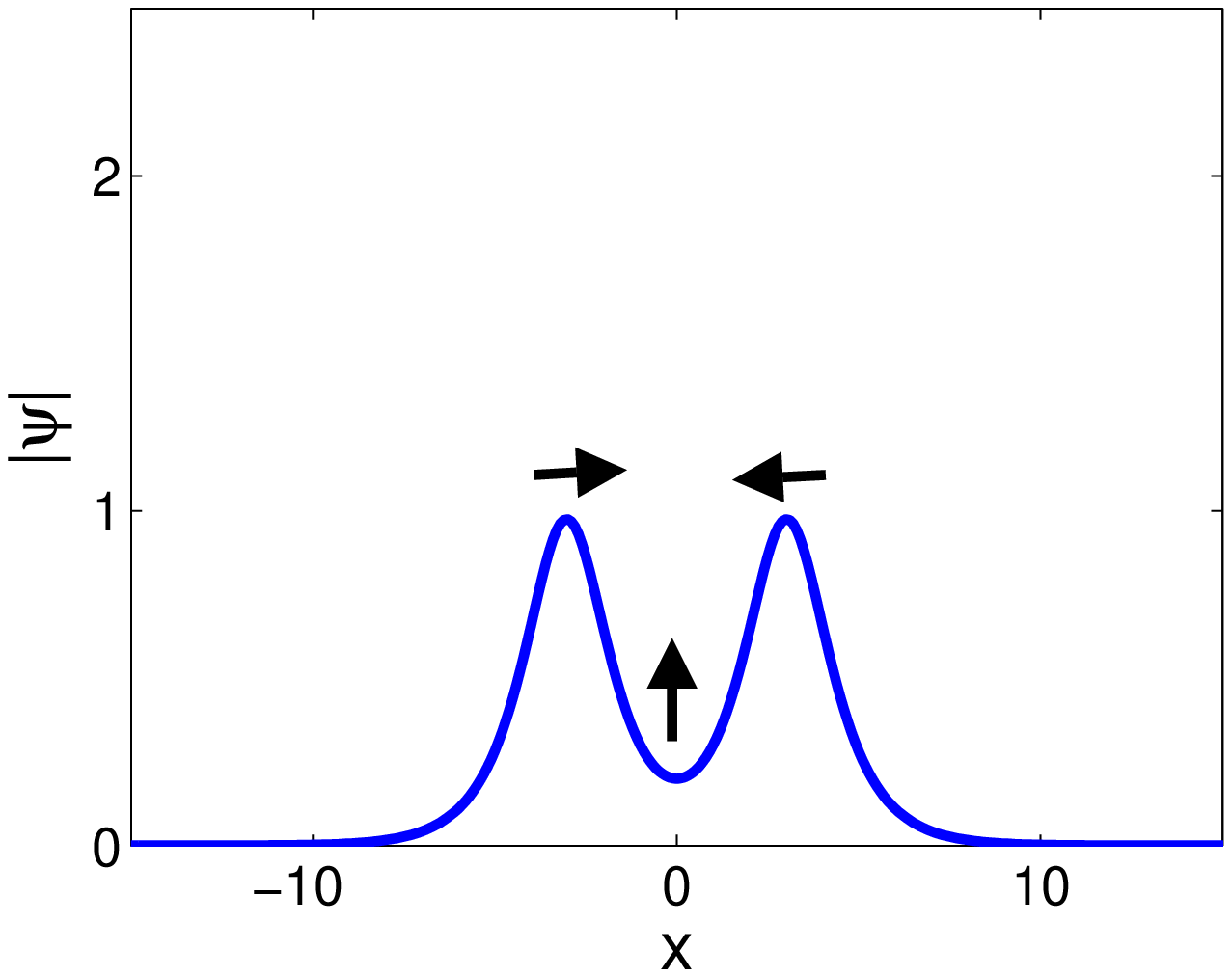}
\par\end{centering}

\begin{centering}
(c)\qquad{}\qquad{}\qquad{}\qquad{}\qquad{}\qquad{}\qquad{}~~~~~~~~~~(d)
\par\end{centering}

\caption{\emph{A breather with large relative distance associated with ZS eigenvalues
$\zeta_{1}=0.75i,$ $\zeta_{2}=0.25i$ are shown at (a) $t=\pi$ and
(b) $t=2\pi$. A breather with small relative distance associated
with $\zeta_{1}=0.55i$ and $\zeta_{2}=0.45i$ are shown at (c) $t=5\pi$
and (d) $t=10\pi$ }\label{fig:Breather large relative distance}}
\end{figure}

Two purely imaginary ZS eigenvalues are associated with quiescent
breathers. This corresponds to two solitons with zero velocity. One
could think of this as a nonlinear analogue of a superposition of
two solitons. Breathers associated with ZS eigenvalues $\zeta_{1,2}=i\eta_{1,2}$
can be reconstructed from the following exact solution \cite{key-38}
\begin{equation}
q\left(x,t\right)=\frac{4i\left(\eta_{2}^{2}-\eta_{1}^{2}\right)\left[\eta_{1}\mbox{cosh}\left(2\eta_{2}x\right)e^{2i\eta_{1}^{2}t}-\eta_{2}\mbox{cosh}\left(2\eta_{1}x\right)e^{2i\eta_{2}^{2}t}\right]}{\left(\eta_{2}-\eta_{1}\right)^{2}C_{+}+\left(\eta_{1}+\eta_{2}\right)^{2}C_{-}-4\eta_{1}\eta_{2}\mbox{cos}\phi},\label{eq:breather}\end{equation}
where $\eta_{2}>\eta_{1}$, $C_{\pm}=\mbox{cosh}\left[2\left(\eta_{2}\pm\eta_{1}\right)x\right]$
and $\phi=2\left(\eta_{2}^{2}-\eta_{1}^{2}\right)t$. The temporal
period of these solutions is $T=\pi/\left(\eta_{2}^{2}-\eta_{1}^{2}\right)$. 

Quiescent breathers are time periodic solutions that oscillate between
two states, similar to a pendulum swinging to and fro from one side
to the other. The first state consists of a thin soliton with amplitude
$A=2\left(\eta_{1}+\eta\right).$ This soliton is sandwiched between
two smaller lateral waves. This state appears independently from the
choice of $\eta_{1}$ and $\eta_{2}.$ The second state of the breather
depends on the relative distance between the ZS eigenvalues, defined
as\[
d_{rel}=\frac{\eta_{1}-\eta_{2}}{\eta_{1}}.\]
For large relative distances ($d_{rel}\geq0.5$) the second state
consists of a single bell-shaped soliton with a large width. For small
relative distances ($d_{rel}\ll1$ ) the second state consists of
two separate solitons. The differences are illustrated in Figure~\ref{fig:Breather large relative distance}.
In Figure~\ref{fig:Breather large relative distance}~(a) and (b)
we show the two different states that arise for a large relative distance
where $\eta_{1}=0.75$ and $\eta_{2}=0.25$. In Figure~\ref{fig:Breather large relative distance}~(c)
and (d) we show the two states associated with a small relative distance
where $\eta_{1}=0.55$ and $\eta_{2}=0.45$. Note that in both cases
the first state is similar, whereas the second states are rather different.

For soliton attractors of the PDNLS equation two purely imaginary
ZS eigenvalues are observed in Type~III and IV attractors. This breather-like
state, associated with two purely imaginary ZS eigenvalues, are associated
with the first state of the breathers discussed above, i.e. a state
consisting of a tall thin soliton with two lateral waves. Examples
of this state is shown in Figure~\ref{eq:Two complex ZS eigenvalues}~(a)
and (c).

\subsubsection*{Two complex ZS eigenvalues symmetric about the imaginary axis}

\begin{figure}[t]
\begin{centering}
\includegraphics[scale=0.4]{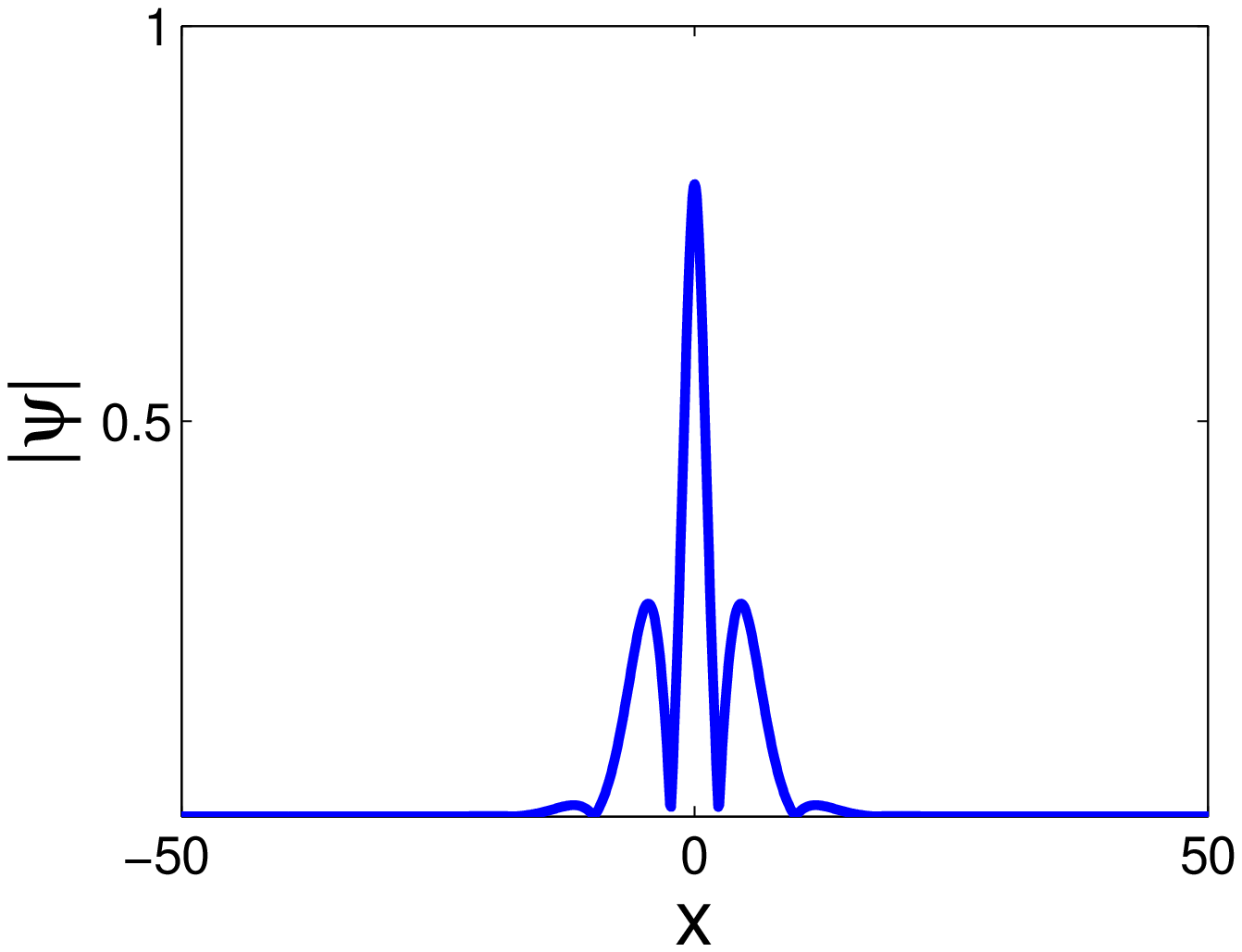}~~~~~\includegraphics[scale=0.4]{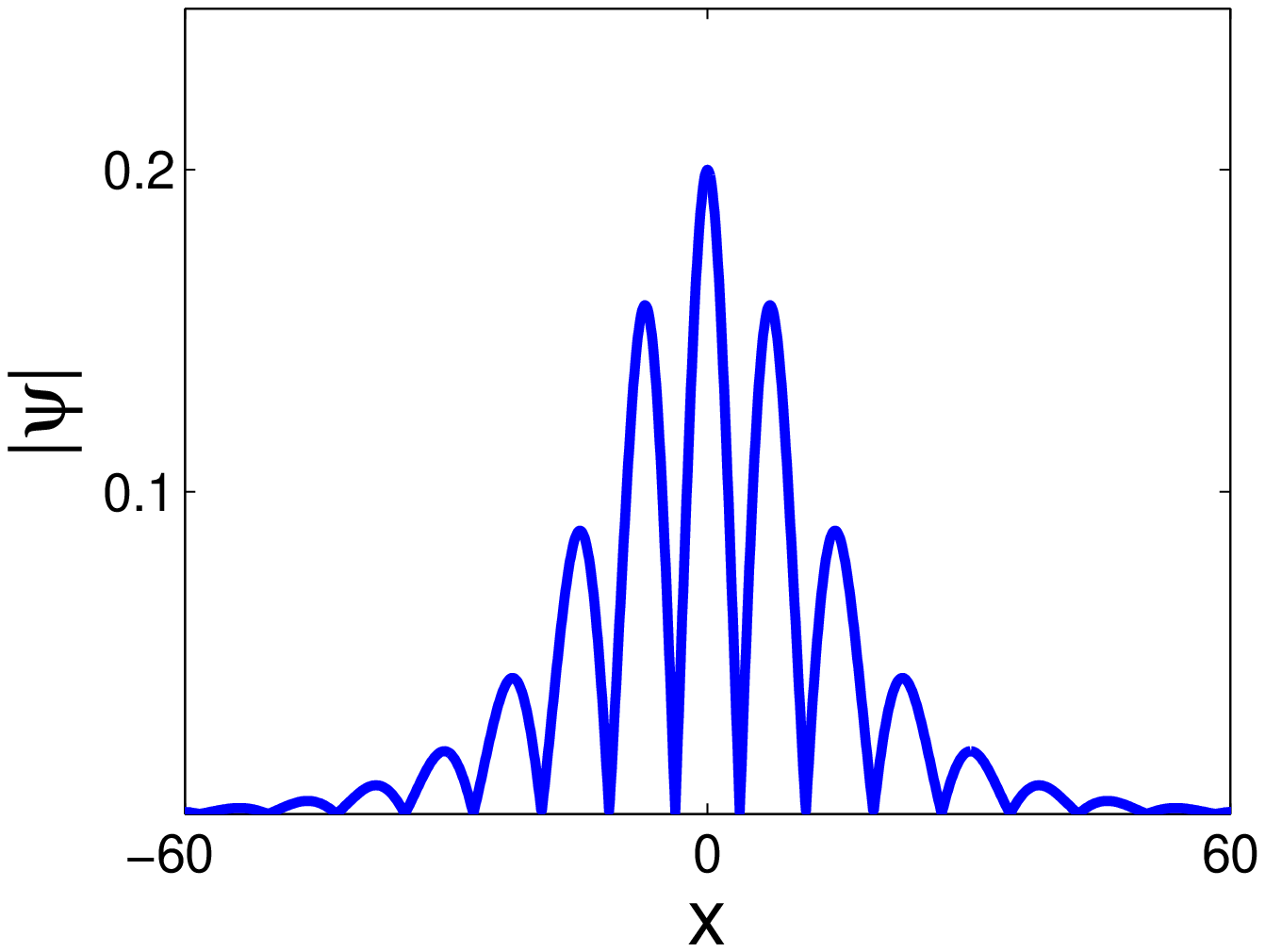}
\par\end{centering}

\begin{centering}
(a)\qquad{}\qquad{}\qquad{}\qquad{}\qquad{}\qquad{}~~~~~~~~~~~~~(b)
\par\end{centering}

\caption{\emph{Climax of collision associated with ZS eigenvalues (a) $\zeta=\pm0.2+0.2i$}
\emph{and (b) $\zeta=0.2\pm0.05i$.\label{fig:Collision}}}
\end{figure}

Two complex ZS eigenvalues \begin{equation}
\zeta=\pm\xi+i\eta\label{eq:Two complex ZS eigenvalues}\end{equation}
 are associated with two solitons with equal amplitudes travelling
in opposite directions with velocities $V=\pm4\xi$. It is well known
that the collision of two solitons in the NLS equation is reflectionless,
resulting in no alteration to the solitons after the collision, except
for a change in phase. The reconstructed solution associated with
two ZS eigenvalues \eqref{eq:Two complex ZS eigenvalues} are given
by \cite{key-38}\begin{equation}
q\left(x,t\right)=-8i\eta\xi\,\frac{A+iB}{D}\, e^{-2i\left(\eta^{2}+\xi^{2}\right)t},\label{eq:Soliton collision}\end{equation}
where\[
A=4\mbox{cosh}4\eta\xi t\,\left[\xi\mbox{cosh}2\eta x\:\mbox{cos}2\xi x-\eta\mbox{sinh}2\eta x\:\mbox{sin}2\xi x\right],\]
\[
B=\mbox{sinh}\left(4\eta\xi t\right)\left[\xi\mbox{sinh}2\eta x\:\mbox{sin}2\xi x+\eta\mbox{cosh}2\eta x\:\mbox{cos}\left(2\xi x\right)\right],\]
and\[
D=\xi^{2}\mbox{cosh}4\eta x+\left(\eta^{2}+\xi^{2}\right)\mbox{cosh}8\eta\xi t-\eta^{2}\mbox{cos}4\xi x.\]
We refer to the solution \eqref{eq:Soliton collision} at $t=0$ as
the climax of collision. This is the point where the two solitons
are least recognisable. 

The behaviour of the climax of the collision depends on the amplitude/velocity
(a/v) ratio defined as \begin{equation}
r_{col}=\eta/\xi.\label{eq:amplitude/velocity ratio}\end{equation}
For large (a/v) ratios $r_{col}>0.5$ the climax of collision resembles
the breather-like state consisting of a tall thin soliton sandwiched
between two lateral waves. For soliton attractors of the PDNLS equation
we interpret this as a breather-like structure. In contrast, for $r_{col}\ll1$
the climax of collision consists of a central soliton with multiple
lateral waves on each side. Figure~\ref{fig:Collision} shows the
difference between these cases. In Figure~\eqref{fig:Collision}~(a)
the climax of collision is shown for $\eta=0.2$ and $\xi=0.2$, corresponding
to a large a/v ratio. The solution consists of a tall soliton surrounded
by two lateral waves. In Figure~\eqref{fig:Collision}~(b) the climax
of collision is shown for $\eta=0.05$ and $\xi=0.2$ , corresponding
to a small a/v ratio. Here we see multiple lateral waves beside the
central soliton.

For soliton attractors of the PDNLS equation we associate symmetric
ZS eigenvalues large a/v ratios with the lateral wave breather-like
structures, similar to those associated with two purely imaginary
ZS eigenvalues. Small a/v ratios are interpreted as radiation tail
structures, such as those observed in Type~II attractors. For splitting
eigenvalues of Type~IV attractors, the a/v ratio is initially large,
due to the smallness of the real part. During this time the soliton
content reveals a breather-like structure, consisting of a soliton
and two lateral waves. As the splitting eigenvalues approach the real
axis, the a/v ratio approaches zero. This shows that the breather-like
structure is broken. The result is that the lateral waves decouple from the
central wave to form radiation waves.

\end{document}